\documentstyle[epsfig,amsmath,amssymb,graphicx,soul,color,subfigure,lscape]{jaa}
\begin{document}
\title[Kelvin--Helmholtz Instability in Solar Atmosphere]{On Modeling the Kelvin--Helmholtz Instability in Solar Atmosphere}

\author[Zhelyazkov]{I.~Zhelyazkov \\ \\
Faculty of Physics, Sofia University,
1164 Sofia, Bulgaria \\ \\
}
\pubyear{2014--2015}
\volume{xx}
\date{Received October 30, 2014; accepted February 20, 2015}
\maketitle
\label{firstpage}

\begin{abstract}
In the present review article, we discuss the recent developments in studying the Kelvin--Helmholtz (KH) instability of magnetohydrodynamic (MHD) waves propagating in various solar magnetic structures.  The main description is on the modeling of KH instability developing in the coronal mass ejections (CMEs), and contributes to the triggering of wave turbulence subsequently leading to the coronal heating.  KH instability of MHD waves in coronal active regions recently observed and imaged in unprecedented detail in EUV high cadence, high-resolution observations by \emph{SDO}/AIA, and spectroscopic observations by \emph{Hinode}/EIS instrument, is posing now challenge for its realistic modeling.  It is shown that considering the solar mass flows of CMEs as moving cylindrical twisted magnetic flux tubes, the observed instability can be explained in terms of unstable $m = -3$ MHD mode.  We also describe the occurrence of the KH instability in solar jets.  The obtained critical jet speeds for the instability onset as well as the linear wave growth rates are in good agreement with the observational data of solar jets.
\end{abstract}


\section{Introduction}
\label{sec:intro}
One of the biggest questions about the solar corona is its heating mechanism. The corona is a thousand times hotter than the Sun's visible surface.  However, its heating mechanisms are still not well-understood.  Scientists have suggested that MHD waves propagating on flowing magnetic flux tubes can become unstable as the flow speed exceeds some critical value, and the developing instability (usually the Kelvin--Helmholtz) might trigger turbulence which causes heating.  Magnetic flux tubes existing in the photosphere and chromosphere of the Sun are considered to be narrow bundles of strong magnetic field lines that rapidly expand with height in the solar atmosphere (\emph{e.g.}, Solanki 1993). The fundamental modes of linear oscillations of these flux tubes are typically identified with longitudinal, transverse, and torsional tubewaves (\emph{e.g.}, Defouw 1976; Roberts \& Webb 1978; Roberts 1979, 1981, 1991; Spruit 1981; Edwin \& Roberts 1983; Hollweg 1985; Roberts \& Ulmschneider 1997; Priest 2014).

\subsection{Observational signature of kink waves in the solar corona}
\label{subsec:observations}

Transverse, or kink, oscillations of solar coronal loops are among the most often studied dynamical phenomena in the corona since their discovery (Aschwanden \emph{et al.}\ 1999; Nakariakov \emph{et al.}\ 1999) with the \emph{Transition Region and Coronal Explorer\/} (\emph{TRACE}; Handy \emph{et al.}\ 1999).  Wang \& Solanki (2004) reported the first evidence for vertical kink oscillations on a loop observed by \emph{TRACE\/} in the 195~\AA\ bandpass at the solar limb.  Kukhianidze, Zaqarashvili \& Khutsishvili (2006) reported spectroscopic observation of kink waves in solar limb spicules with the $53$ cm coronagraph of the Abastumani Astrophysical Observatory.  Li \& Gan (2006) detected an oscillatory shrinkage in a \emph{TRACE\/} $195$~\AA\ loop during a flare impulse phase and showed that the loop shrinkage is in the form of an oscillation, with a period of about $150$~s and an amplitude of about $300$~km.  Using temporal series image data from the Coronal Diagnostic Spectrometer (CDS) on \emph{SOHO}, O'Shea \emph{et al.}\ (2007) measured all statistically significant frequencies present in oscillations found in flaring active region loops at the temperature of O {\footnotesize\textsc{v}} $629$ line.  By measuring the distances traveled by
three propagating disturbances and by calculating their propagation speeds, the authors found evidence that standing fast kink waves were present in flaring cool transition region loops.  Recently, the ubiquitous presence of kink waves are found in various coronal structures using high-resolution observations by \emph{SDO}/AIA (Aschwanden \& Schrijver 2011; White, Verwichte \& Foullon 2012; Srivastava \& Goossens 2013; Yang \emph{et al.}\ 2013).

Observational evidence for the existence of transverse MHD waves in different regions of the solar atmosphere was given by high resolution observations performed by the Solar Optical Telescope (SOT) and the X-Ray Telescope (XRT) on board the \emph{Hinode\/} Solar Observatory.  According to De Pontieu \emph{et al.}\ (2007b) and Cirtain \emph{et al.}\ (2007), signatures of Alfv\'en waves were observed by the SOT and XRT instruments, respectively.  Moreover, Alfv\'en waves were also reported by Tomczyk \emph{et al.}\ (2007), who used the Coronal Multi-Channel Polarimeter of the US National Solar Observatory.  Interpretations of these observations were given by Van Doorsselaere, Nakariakov \& Verwichte (2008) and Antolin \emph{et al.}\ (2009), who concluded that the reported observational results describe kink waves.  On the other hand, Goossens \emph{et al.}\ (2012) argue that the fundamental radial modes of kink ($m = 1$) waves with phase velocity between the internal and external Alfv\'en velocities can be considered as surface Alfv\'en waves (or Alfv\'enic waves in the nomenclature of Goossens \emph{et al.}\ 2009).  In other words, the transverse waves as observed in the solar corona by Tomczyk \emph{et al.}\ (2007), De Pontieu \emph{et al.}\ (2007b), and Cirtain \emph{et al.}\ (2007) can be considered as surface Alfv\'en waves.  Alfv\'enic oscillations were confidently detected by McIntosh \emph{et al.}\ (2011), who reported indirect evidence for such waves found in observations with high-resolution extreme ultra-violet (EUV) imagers, such as the Atmospheric Imaging Assembly (AIA; Lemen \emph{et al.}\ 2012) on board the \emph{Solar Dynamics Observatory\/} (\emph{SDO}; Pesnell, Thompson \& Chamberlin 2012).  Spectroscopic signatures of Alfv\'enic waves was also established by Tian \emph{et al.}\ (2012) by studying persistent Doppler shift oscillations observed with \emph{Hinode}/EIS in the solar corona.  Since the naming of transverse oscillations observed in solar structures remains highly controversial, with definitions revolving around ``Alfv\'en,'' ``Alfv\'enic,'' and ``magnetosonic kink'' terminology, we will deliberately choose to describe the observed periodic motions simply as transverse kink waves.  This is currently the most unopposed description of such wave phenomena in the solar atmosphere.

\subsection{Coronal seismology}
\label{subsec:seismology}
MHD waves and oscillations being ubiquitous in the solar corona. Using the technique of coronal seismology (for a review see Nakariakov \& Verwichte 2005), they allow us to measure the coronal magnetic field and loop structuring.  Verwichte \emph{et al.}\ (2009) performed seismology of a large solar coronal loop from EUV imager (on board the \emph{STEREO}) observation of its  transverse oscillation.  The three-dimensional loop geometry was determined using a three-dimensional reconstruction with a semicircular loop model, which allows for an accurate measurement of the loop length.  The plane of wave
polarization was found from comparison with a simulated loop model and showed that the oscillation is a fundamental horizontally polarized fast kink mode.  A detailed analysis of coronal kink mode loop oscillations with AIA/\emph{SDO} performed by Aschwanden \& Schrijver (2011) allowed them to determine the exact footpoint locations and loop length with stereoscopic triangulation using \emph{STEREO}/EUVI/A data.  They also modeled the magnetic field in the oscillating loop using Helioseismic and Magnetic Imager/\emph{SDO\/} magnetogram data and a potential-field model and found agreement with the seismological value of the magnetic field, $B_{\rm kink} = 4.0 \pm 0.7$~G.  Liu \emph{et al.}\ (2012) presented the first unambiguous detection of quasi-periodic wave trains within the broad pulse of a global EUV wave (so-called EIT wave) occurring on the limb and sequential transfer oscillations detected by \emph{SDO}/AIA.  These observations provided compelling evidence of the fast-mode MHD wave nature of the global EUV wave.  The first evidence of transverse oscillations of a multistranded loop with growing amplitudes and internal coupling observed by the AIA/\emph{SDO\/} was obtained by Wang \emph{et al.}\ (2012).  The loop oscillation event occurred on 2011 March 8, was triggered by a coronal mass ejection (CME).  The multiwavelength analysis revealed the presence of multithermal strands in the oscillating loop, whose dynamic behaviors were temperature-dependent, showing differences in their oscillation amplitudes, phases, and emission evolution.  The authors suggested that the amplitude-growing kink oscillations may be a result of continuous non-periodic driving by magnetic deformation of the CME, which deposits energy into the loop system at a rate faster than its loss.  White \emph{et al.}\ (2012) reported and analyzed the observation of a vertically polarized transverse oscillation in a hot coronal loop with the AIA/\emph{SDO}, following a linked coronal-flare mass-ejection event on the 3 November 2010.  The oscillating coronal loop is observed off the east solar limb and exclusively in the $131$~\AA\ and $94$~\AA\ bandpasses, indicating a loop plasma of temperature in the range of $9$--$11$~MK.  They also presented the periods and damping times of the second and third harmonic of the transverse kink oscillation.  Decaying and decayless transverse kink oscillations of a coronal loop were also observed by Nistic\`o, Nakariakov \& Verwichte (2013) in an active region with AIA/\emph{SDO\/} before and after a flare.  The authors showed that before and well after the occurrence of the flare, the loops experience low-amplitude decayless oscillations.  The natural kink modes excited in the loops are decaying oscillations. It was modeled recently that any external driver, for example, EUV wave, CME, \emph{etc.}, may trigger the decayless forced oscillations in coronal loops (Murawski \emph{et al.}\ 2015).  Srivastava \& Goossens (2013) again on using AIA/\emph{SDO\/} observed X6.9-class flare-induced vertical kink oscillations with periods of $795$~s and $530$~s in a large-scale plasma curtain.  On the magnetic surface of the curtain where the density is inhomogeneous due to coronal dimming, non-decaying vertical oscillations were also observed (with a period of ${\approx}763$--$896$~s).  The authors infer that the global large-scale disturbance triggers vertical kink oscillations in the deeper layers as well as on the surface of the large-scale plasma curtain.  Guo \emph{et al.}\ (2015) reported the observation of the first two harmonics of the horizontally polarized kink waves excited in a coronal loop system lying southeast of NOAA AR 11719 on 2013 April 11.  The detected periods of the fundamental mode, $P_1$, its first overtone, $P_2$, in the northern half, and that in the southern one were $530.2 \pm 13.3$, $300.4 \pm 27.7$, and $334.7 \pm 22.1$~s, respectively.  The periods of the first overtone in the two halves were the same considering uncertainties in the measurement.  The authors estimated the average electron density, temperature, and length of the loop system as $(5.1 \pm 0.8) \times 10^8$ cm$^{-3}$, $0.65 \pm 0.06$~MK, and $203.8 \pm 13.8$~Mm, respectively.  As a zeroth-order estimation, the magnetic field strength, $B = 8.2 \pm 1.0$~G, derived by the coronal seismology using the fundamental kink mode matches with that derived by a potential field model.

It is well established that tubular kink modes are the one of natural modes excited in various magnetic waveguides in the solar atmosphere, and they are significant also in diagnosing the physical conditions of the localized corona.  In addition, such waves when excited on the flowing jets or flux ropes (CMEs) may undergo to the instability and further may lead to play a role in local plasma dynamics and turbulent heating.  We will review some recent modeling efforts in this context taking the inference of CME and jets.  Therefore, in next we will briefly review about these two dynamical phenomena in the solar atmosphere.

\subsection{Jets in the solar atmosphere}
\label{subsec:jets}
It is observationally well established that the dynamic solar atmosphere contains different kind of jets.  In the coronal hole regions the ambient magnetic field is nearly vertical and often unipolar, and interacting with the emerging field gives rise to reconnection followed by mass ejections, with collimated hot-plasma flows commonly termed jets (Yokoyama \& Shibata 1995). After the launch of \emph{Hinode}/XRT it was discovered that these jets occur more frequently than previously thought, at a frequency of $60$ jets per day (Savcheva \emph{et al.}\ 2007) and 10 jets per hour (as reported by Cirtain \emph{et al.}\ 2007). Small-scale solar eruptions seen in different wavelengths are often termed differently, for instance, H$\alpha$ surges (Newton 1942), spicules (\emph{e.g.}, Secchi 1877; Beckers 1972, and references therein), type II spicules (De Pontieu \emph{et al.}\ 2007c, 2012), macro-spicules (Bohlin \emph{et al.}\ 1975), mottles (De Pontieu \emph{et al.}\ 2007a), UV jets (Brueckner \& Bartoe 1983), EUV jets (Budnik \emph{et al.}\ 1998), and X-ray jets (Shibata \emph{et al.}\ 1992).  Recent high-resolution and high-cadence observations (\emph{e.g.}, using \emph{SOHO}, \emph{Hinode}, \emph{STEREO}, \emph{SDO}) have allowed a detailed study of coronal jets, providing information on their inherent dynamic behavior (see, \emph{e.g.}, Kamio \emph{et al.}\ 2010; Shen \emph{et al.}\ 2011; Srivastava \& Murawski 2011; Morton, Srivastava \& Erd\'elyi 2012; Pereira, De Pontieu \& Carlsson 2013; Zheng \emph{et al.}\ 2013; Kayshap, Strivastava \& Murawski 2013; Kayshap \emph{et al.}\ 2013; Verwichte \emph{et al.}\ 2013; Zhang \& Ji 2014).  The magnetic reconnection is found to be a major mechanism for the triggering of various solar jets (direct $\vec{j} \times \vec{B}$ force or its indirect consequences; Shibata \emph{et al.}\ 1992; Yokoyama \& Shibata 1995; Nishizuka \emph{et al.}\ 2008; Murawski, Srivastava \& Zaqarashvili 2011; Kayshap, Srivastava \& Murawski 2013).  In spite of this, the various instabilities (Pariat \emph{et al.}\ 2009), MHD waves (Kudoh \& Shibata 1999; Cirtain et al. 2007; Jel\'{i}nek \emph{et al.}\ 2015), MHD pulses/shocks trains (Hansteen \emph{et al.}\ 2006) are found to be the other possible mechanisms for the evolution of various jets in the solar atmosphere.  In the present article, we will also examine the role of KH instability in the solar jets.

\subsection{Coronal mass ejections and magnetic flux ropes}
\label{subsec:CMEs}

Alongside the small-scale solar eruptions there also exist large scale solar mass eruptions called coronal mass ejections (CMEs), which are probably the most important sources of adverse space weather effects (see, \emph{e.g.}, Howard \emph{et al.}\ 2006; Zhang \emph{et al.}\ 2007; Temmer \emph{et al.}\ 2010).  They are often associated with dramatic changes of coronal magnetic fields (\emph{e.g.}, Zhang \& Low 2005; Liu \emph{et al.}\ 2009; Su \& van Ballegooijen 2012).  Coronal mass ejections are huge clouds of magnetized plasma (about $10^{15\mbox{--}16}$~g) that erupt from the solar corona into interplanetary space.  CMEs were observationally discovered 40 years ago (Gosling \emph{et al.}\ 1974)---they were observed with the white light coronagraph experiment on board \emph{Skylab\/} during the first 118 days of the mission.  CMEs propagate in the heliosphere with velocities ranging from $20$ to $3200$~km\,s$^{-1}$ with an average speed of $489$~km\,s$^{-1}$, according to \emph{SOHO}/LASCO coronagraph (Brueckner, Howard \& Koomen 1995) measurements between 1996 and 2003.  The CME on July 23, 2012, captured by \emph{STEREO}-Ahead's Cor2 (coronograph), clocked in between $2900$ and $3540$ kilometers per second and it is the fastest one ever seen by \emph{STEREO}.  Depending on the velocity magnitude, according to the CME SCORE Scale (Evans \emph{et al.}\ 2013), one can distinguish five categories of CMEs, notably: S-type (${<}500$~km\,s$^{-1}$), C-type (Common) ($500$--$999$~km\,s$^{-1}$), O-type (Occasional) ($1000$--$1999$~km\,s$^{-1}$), R-type (Rare) ($2000$--$2999$~km\,s$^{-1}$), and ER-type (Extremely Rare) (${>}3000$~km\,s$^{-1}$).  The physical mechanisms that initiate and drive solar eruptions were discussed in many papers over past four decades (see, \emph{e.g.}, Chandra \emph{et al.}\ 2014; Schmieder, D\'emoulin \& Aulanier 2013; Chandra \emph{et al.}\ 2011; Aulanier \emph{et al.}\ 2010; Forbes \emph{et al.}\ 2006; T\"{o}r\"{o}k \& Kliem 2005, and references therein). Very recently, according to Aulanier (2014) ``no more than two distinct physical mechanisms can actually initiate and drive prominence eruptions: the \emph{magnetic breakout\/} and the \emph{torus instability}.  In this view, all other processes (including flux emergence, flux cancellation, flare reconnection and long-range couplings) should be considered as various ways that lead to, or that strengthen, one of the aforementioned driving mechanisms.''  Studies using the data sets from (among others) the \emph{SOHO}, \emph{TRACE}, \emph{Wind}, \emph{ACE}, \emph{STEREO}, and \emph{SDO} spacecraft, along with ground-based instruments, have improved our knowledge of the origins and development of CMEs at the Sun (Webb \& Howard 2012; Joshi \emph{et al.}\ 2013a,b; Landi \& Miralles 2014).

  \begin{figure}[ht]
    \centering
    \includegraphics[height=0.247\textheight]{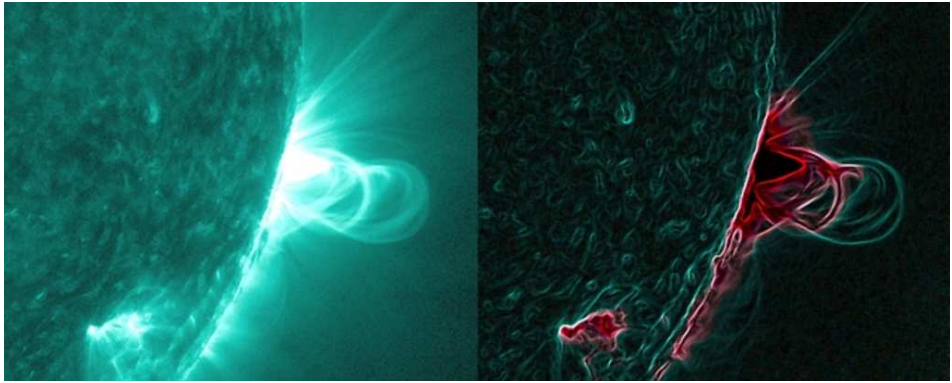}
    \caption{The image on the left shows a series of magnetic loops on the Sun, as captured by NASA's \emph{SDO\/} on July 18, 2012.  The image on the right has been processed to highlight the edges of each loop and make the structure more clear. A series of loops such as this is known as a flux rope, and these lie at the heart of eruptions on the Sun known as CMEs. This is the first time scientists were able to discern the timing of a flux rope's formation.  \emph{Credit to: NASA/}SDO.}
    \label{fig:fig1}
  \end{figure}
It is generally accepted that CMEs are the results of eruptions of magnetic flux ropes (MFRs) (see Fig.~\ref{fig:fig1}).  However, there is heated debate on whether MFRs exist prior to the eruptions or if they are formed during the eruptions.  Several coronal signatures, for example, filaments, coronal cavities, sigmoid structures, and hot channels (or hot blobs), are proposed as MFRs and observed before the eruption, which support the pre-existing MFR scenario (see, \emph{e.g.}, Cheng \emph{et al.}\ 2011;
Zhang, Cheng \& Ding 2012; Patsourakos, Vourlidas \& Stenborg 2013; Vourlidas 2014; Green \& Kliem 2014; Cheng \emph{et al.}\ 2014; Chen, Bastian \& Gary 2014; Howard \& DeForest 2014; Vemareddy \& Zhang 2014).  According to the second scenario, a new flux rope forms as a result of the reconnection of the magnetic lines of an arcade (a group of arches of field lines) during the
eruption itself.  Observational support for this mechanism was recently reported by Song \emph{et al.}\ (2014).  The authors present an intriguing observation of a solar eruptive event that occurred on 2013 November 21 with the AIA/\emph{SDO}, which shows the formation process of the MFR during the eruption in detail.  After all, the majority of observations support the view that, in at least some CMEs, flux rope formation occurs before launch.  Thus, we are convinced to accept the expanded definition of the CME suggested in Vourlidas \emph{et al.}\ (2013): ``A CME is the eruption of a coherent magnetic, twist-carrying coronal structure with angular width of at least $40^{\circ}$ and able to reach beyond $10$~$R_{\odot}$ which occurs on a time scale of a few minutes to several hours.''

\subsection{Kelvin--Helmholtz instability in untwisted solar jets}
\label{subsec:KH-instability}

It is well-known that Kelvin--Helmholtz instabilities occur when two fluids of different densities or different speeds flow onto each other. In the solar atmosphere, which is made of a very hot and practically fully ionized plasma, the two flows come from an expanse of plasma erupting off the Sun's surface as it passes by plasma that is not erupting.  The difference in flow speeds and densities across this boundary sparks the instability that builds into the waves.  When the instability reaches its nonlinear stage, vortices might form, reconnection might be initiated and plasma structures might detach.  A concise but very good exploration of KH instabilities in the solar atmosphere in view of their interpretation from observations we can find in Taroyan \& Ruderman (2011).  A general review of the effects of fundamental parameters like magnetic field, shear velocity and wavenumber on the growth rate of the magnetic KH instability occurring in the solar atmosphere have been presented by Cavus \& Kazkapan (2013).  Considering two semi-infinite flowing magnetized plasmas, they studied the conditions for occurrence and non-occurrence of instability that are determined for the different values of magnetic field, shear velocity and wavenumber in the solar atmosphere.  The authors obtained the critical values of shear velocity of $600$--$900$~km\,s$^{-1}$ and magnetic field of $4$--$10$~G, which are consistent with observations.  KH instability and resonant flow instability for a coronal plume in more realistic geometries (slab and cylinder) have been studied by Andries \& Goossens (2001).  They obtained an analytical inequality for the occurrence of KH instability and showed that the instability that will most probably occur in coronal plumes is due to an Alfv\'en resonance of slow psedosurface (body) MHD modes.  KH instability of kink waves in soft X-ray coronal jets and solar spicules (in cylindrical geometry) were investigated by Vasheghani Farahani \emph{et al.}\ (2009) and Zhelyazkov (2012a), respectively.  It was shown in both cases that kink modes are stable against the KH instability because the critical jet's speeds turned out to be far beyond the velocities accessible for considered jets---in particular, for the soft X-ray coronal jet that velocity must be at least $3576$~km\,s$^{-1}$, and $882$~km\,s$^{-1}$ for spicules.  The critical speed of Vasheghani Farahani \emph{et al.}\ (2009) was numerically confirmed and refined by Zhelyazkov (2012b) (see also Zhelyazkov 2013 and references therein)---the updated critical speed is $3448$~km\,s$^{-1}$.  A more careful evaluation yields $3200$~km\,s$^{-1}$, that velocity lies at the upper limit of soft X-ray coronal jet speeds (Shimojo \& Shibata 2000; Madjarska 2011).

\subsection{Waves and instabilities in twisted solar magnetic structures}
\label{subsec:twisted}

A major step in studying the waves and instabilities in magnetically structured solar atmosphere is the consideration of a twist of the background magnetic field.  Twisted magnetic flux tubes have been investigated for many years primarily in the context of tube stability or in relation to the MHD wave resonant absorption.  Magnetic tubes are subject to the kink instability when
the twist exceeds a critical value (Lundquist 1951; Hood \& Priest 1979).  Oscillations and waves and their stability in twisted magnetic flux tubes without flow have been studied in the framework of the normal mode analysis in earlier works (see Dungey
\& Loughhead 1954; Roberts 1956; Trehan \& Reid 1958; Bogdan 1984; Bennett, Roberts \& Narain 1999; Erd\'elyi \& Carter 2006; Erd\'elyi \& Fedun 2006; Erd\'elyi \& Fedun 2007; Erd\'elyi \& Fedun 2010; Ruderman 2007; Carter \& Erd\'elyi 2008; Ruderman \& Erd\'elyi 2009).  Most of these papers deal with relatively simple twisted magnetic configurations: incompressible plasma cylinders/slabs surrounded by perfectly conducting unmagnetized plasma or a medium with an untwisted homogeneous magnetic field. Erd\'elyi \& Fedun (2006) were the first to study the wave propagation in a twisted cylindrical magnetic flux tube embedded in an incompressible but also magnetically twisted plasma.  In the series of papers by Erd\'elyi and collaborators listed above there have been studied the dispersion relations of kink ($m = 1$) and sausage ($m = 0$) MHD modes in various more complex geometries and magnetic field topologies (for details, see Zhelyazkov \& Zaqarashvili 2012).  The only work studying the wave propagation in a twisted magnetic tube with a mass density variation along the tube is that of Ruderman (2007).  With an asymptotic analysis, he showed that the eigenmodes and the eigenfrequencies of the kink and fluting-like oscillations are described by a classical
Sturm--Liouville problem for a second-order ordinary differential equation.  Transverse oscillations of coronal loops have been studied by Ruderman \& Erd\'elyi (2009) who explored the effects of stratification, loop expansion, loop curvature, non-circular cross-section, loop shape, and magnetic twist on the damping of kink waves due to resonant absorption.  An extended review of Alfv\'en waves in the solar atmosphere (both in untwisted and twisted flux tubes) was recently presented by Mathioudakis, Jess \& Erd\'elyi (2013).

One important question is how a flow along a twisted magnetic flux tube will change the dispersion properties of the propagating
modes and their stability.  It turns out that the flow may decrease the threshold for the kink instability, as was tested experimentally in a laboratory twisted plasma column (Furno \emph{et al.}\ 2007).  This observation was theoretically confirmed by Zaqarashvili \emph{et al.}\ (2010). The authors studied the influence of axial mass flows on the stability of an isolated twisted magnetic tube of incompressible plasma embedded in a perfectly conducting unmagnetized plasma.  Two main results were found. First, the axial mass flow reduces the threshold of the kink instability in twisted magnetic tubes. Second, the twist of the magnetic field leads to the KH instability of sub-Alfv\'enic flows for the harmonics with a sufficiently large azimuthal mode number $m$.  D\'{i}az \emph{et al.}\ (2011) also studied the equilibrium and stability of twisted magnetic flux tubes with mass flows, but for flows along the field lines. The authors focused on the stability and oscillatory modes of magnetic tubes with a uniform twist in a zero-beta plasma surrounded by a uniform cold plasma embedded in a purely longitudinal magnetic filed. Regarding the equilibrium, the authors claimed that the only value of the flow that satisfies the equations for their magnetic field configuration is a super-Alfv\'enic one.  The main conclusion is that the twisted tube is subject to the kink instability unless the magnetic field pitch is very high, since the Lundquist criterion is significantly lowered. This is caused by the requirement of having an Alfv\'en Mach number greater than $1$, so the magnetic pressure balances the magnetic field tension and fluid inertia.  The authors suggest that this type of instability might be observed in some solar atmospheric structures, such as surges.  Soler \emph{et al.}\ (2010) in the zero-beta approximation, in cylindrical geometry, investigated the stability of azimuthal shear flow with a sharp jump of the velocity at the cylinder boundary.  They obtained an analytical expression for the dispersion relation of the unstable MHD modes and found that fluting-like modes can develop a KH instability in timescales comparable to the period of kink oscillations of the flux tube.  The KH instability growth rates increase with the magnitude of the azimuthal wavenumber and decrease with the longitudinal wavenumber.  However, the presence of a small azimuthal component of the magnetic field can suppress the KH instability.  In studying KH instability of kink waves in photospheric twisted flux tubes Zhelyazkov \& Zaqarashvili (2012) have shown that the stability of the waves depends upon four parameters, the density contrast between the flux tube and its environment, the ratio of the background magnetic fields in the two media, the twist of the magnetic field lines inside the tube, and the value of the Alfv\'en Mach number (the ratio of the jet velocity to Alfv\'en speed inside the flux tube).  They assumed that the azimuthal component of the magnetic field in the tube is proportional to the distance from the tube axis and that the tube is only weakly twisted (that is, the ratio of the azimuthal and axial components of the magnetic field is low).  It was obtained that for an isolated twisted photospheric flux tube (magnetically free environment) with density contrast of $2$ (the ratio of the surrounding plasma density to that of the tube itself), magnetic field twist of $0.4$, and Alfv\'en speed $v_{\rm A} = 10$~km\,s$^{-1}$, can trigger an instability of the KH type of $m = 1$ mode.  Any none-zero environment magnetic field slightly increases that critical jet speed for instability onset.  Soler \emph{et al.}\ (2012) have shown that ion--neutral collisions may play a relevant role for the growth rate and evolution of the KH instability in solar partially ionized plasmas such as in, for instance, solar prominences.  They investigated the linear phase of the KH instability at an interface between two partially ionized magnetized plasmas in the presence of a shear flow and found that in the incompressible case, the KH instability is present for any velocity shear regardless of the value of the collision frequency.  In the compressible case, the domain of instability depends strongly on the plasma parameters, especially on the collision frequency and the density contrast.  For high-collision frequencies and low-density contrasts the KH instability is present for super-Alfv\'enic velocity shear only.  For high-density contrasts the threshold velocity shear can be reduced to sub-Alfv\'enic values.  For the particular case of turbulent plumes in prominences, Soler \emph{et al.}\ (2012) concluded that sub-Alfv\'enic flow velocities can trigger the KH instability thanks to the ion--neutral coupling.  Zaqarashvili, V\"{o}r\"{o}s \& Zhelyazkov (2014a) in exploring the KH instability of twisted cylindrical magnetic flux tubes in the solar wind have obtained that twisted magnetic flux tubes can be unstable to KH instability when they move with regard to the solar wind stream.  It was found also that the external axial magnetic field stabilizes KH instability, therefore, the tubes moving along Parker spiral are unstable only for super-Alfv\'nic motions.  However, even a slight twist in the external magnetic field leads to KH instability for any sub-Alfv\'enic motion.  It was established that the unstable harmonics satisfy the relation $\vec{k} \cdot \vec{B} \approx 0$, which corresponds to pure vortices in the incompressible MHD.

After this review of studies on MHD wave characteristics and stability/instability status in various solar atmosphere jets, we shall consider in the next section the modeling KH instability in CMEs because the instability was successfully detected and imaged over last three years namely in CMEs by using the most advanced instruments on board \emph{SDO\/} and \emph{STEREO}.  A summary of the essential findings in this modeling and an outlook for future research on KH instability in various solar atmosphere jets are contained in the last section of the review paper.

\section{Kelvin--Helmholtz instability in coronal mass ejections}
\label{sec:cme}
\begin{figure}[ht]	
\centering
    \includegraphics[height=0.325\textheight]{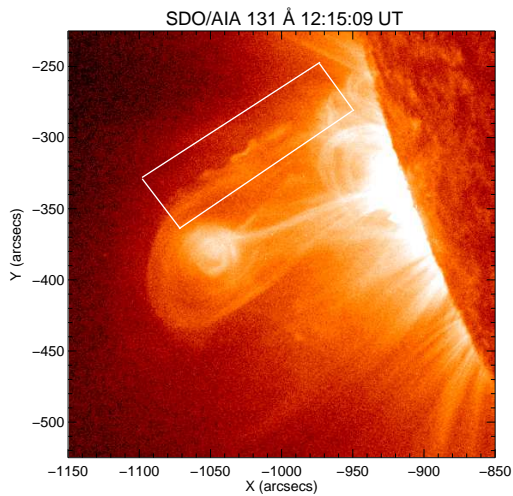}
    \caption{\emph{SDO}/AIA image showing a fast coronal mass ejecta erupting from the Sun, with KH vortices detected on its northern flank closed by rectangle. \emph{Credit to: NASA/}SDO \emph{and R.~Chandra}.}
\label{fig:fig2}
\end{figure}
The first observations of the temporally and spatially resolved evolution of the magnetic KH instability in the solar corona based on unprecedented high-resolution imaging observations of vortices developing at the surface of a fast coronal mass ejecta (less than $150$ Mm above the solar surface in the inner corona) taken with the Atmospheric Imaging Assembly (AIA) on board the \emph{SDO} were reported by Foullon \emph{et al.}\ (2011).  The CME event they studied occurred on 2010 November 3, following a C4.9 \emph{GOES\/} class flare (peaking at 12:15:09 UT from active region NOAA 11121, located near the south-east solar limb).  The instability was only detected in the highest AIA temperature channel, centered on the $131$~\AA~EUV bandpass at $11$~MK.  In this temperature range, the ejection lifting off from the solar surface forms a bubble of enhanced emission against the lower density coronal background (see Fig.~\ref{fig:fig2} as well as Fig.~1 in Foullon \emph{et al.}\ 2011).  Along the northern flank of the ejecta, a train of three to four substructures forms a regular pattern in the intensity contrast.  A similar pattern was reported by Ofman \& Thompson (2011)---the authors presented observations of the formation, propagation, and decay of vortex shaped features in coronal images from the \emph{SDO} associated with an eruption starting at about 2:30 UT on 2010 April 8.  The series of vortices were formed along the interface between an erupting (dimming) region and the surrounding corona.  They ranged in size from several to $10$ arcsec and traveled along the interface at $6$--$14$~km\,s$^{-1}$.  The features were clearly visible in six out of the seven different EUV wavebands of the AIA. Based on the structure, formation, propagation, and decay of these features, Ofman \& Thompson (2011) claim that they identified the event as the first observation of the KH instability in the corona in EUV.  Again, on using the AIA on board the \emph{SDO}, M\"{o}stl, Temmer \& Veronig (2013) observed a S-type coronal mass ejection with an embedded filament on 2011 February 24, revealing quasi-periodic vortex-like structures at the northern side of the filament boundary with a wavelength of approximately $14.4$~Mm and a propagation speed of about $310 \pm 20$~km\,s$^{-1}$.  These structures, according to the authors, could result from the KH instability occurring on the boundary.

An updated and detailed study by Foullon \emph{et al.}\ (2013) of the dynamics and origin of the CME on 2010 November 3 by means of the \emph{Solar TErrestrial RElations Observatory Behind\/} (\emph{STEREO-B}) located eastward of \emph{SDO\/} by $82^{\circ}$ of heliolongitude, and used in conjunction with \emph{SDO\/} give some indication of the magnetic field topology and flow pattern.  At the time of the event, Extreme Ultraviolet Imager (EUVI) from \emph{STEREO}'s Sun--Earth Connection Coronal and Heliospheric Investigation (SECCHI) instrument suite (Howard \emph{et al.}\ 2008) achieved the highest temporal resolution in the $195$~\AA~bandpass: EUVI's images of the active region on the disk were taken every $5$ minutes in this bandpass.  The authors applied differential emission measure (DEM) techniques on the edge of the ejecta to determine the basic plasma parameters---they are summarized in Table~2 of Foullon \emph{et al.}\ 2013.  The main features of the imaged KH instability presented in
\begin{figure}[ht]
   \centering
   \includegraphics[height=.40\textheight]{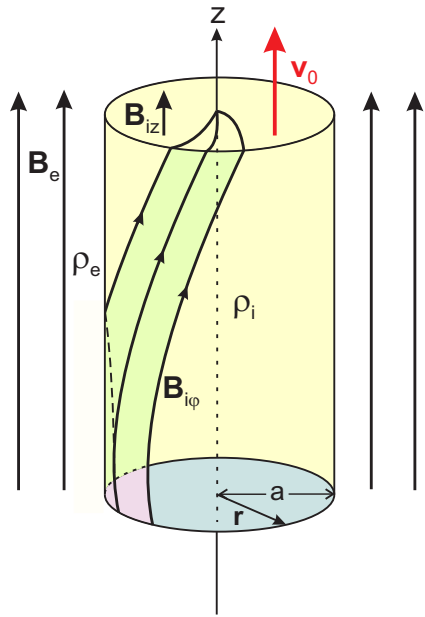}
   \caption{Magnetic field geometry of a coronal mass ejection. \emph{Picture credit to: R.~Erd\'elyi \& V.~Fedun 2010.}}
   \label{fig:fig3}
\end{figure}
Table~3 include (in their notation) the speed of $131$~\AA~CME leading edge, $V_{\rm LE} = 687$~km\,s$^{-1}$, flow shear on the $131$~\AA~CME flank, $V_1 - V_2 = 680 \pm 92$~km\,s$^{-1}$, KH group velocity, $v_{\rm g} = 429 \pm 8$~km\,s$^{-1}$, KH wavelength, $\lambda = 18.5 \pm 0.5$~Mm, and exponential linear growth rate, $\gamma_{\rm KH} = 0.033 \pm 0.012$~s$^{-1}$.  The modeling imaged/registered KH instability by Foullon \emph{et al.}\ (2013) and M\"{o}stl, Temmer \& Veronig (2013) will be done via investigating the stability/instability status of tangential velocity discontinuity at the boundary of the ejecta.

\subsection{Geometry and wave dispersion relation}
\label{subsec:geometry}
As the initial stage of a coronal mass ejection is the formation of a magnetic flux rope, the most appropriate model of the CME imaged by Foullon \emph{et al.}\ (2013) is a twisted magnetic flux tube of radius $a$ (${=}\Delta L/2$, $\Delta L = 4.1$~Mm being the CME's width) and uniform density $\rho_{\rm i}$ embedded in a uniform field environment with density $\rho_{\rm e}$ (see Fig.~\ref{fig:fig3}).  The magnetic field inside the tube is helicoidal with uniform twist, that is, $\vec{B}_{\rm i} = (0, Ar, B_{{\rm i} z})$, where $A$ and $B_{{\rm i} z}$ are constant, while outside it is uniform and directed along the $z$-axis, $\vec{B}_{\rm e} = (0, 0, B_{\rm e})$.  The tube moves along its axis with velocity of $\vec{v}_0$ with regard to the surrounding medium.  The jump of the tangential velocity at the tube boundary then triggers the magnetic KH instability when the jump exceeds a critical value.

Before discussing governing MHD equations and the wave dispersion relation, we specify what kind of plasma each medium is (the moving tube and its environment).  The CME parameters listed in Table~2 in Foullon \emph{et al.}\ 2013 show that the plasma beta inside the flux tube might be equal to $1.5 \pm 1.01$, while that of the cooler coronal plasma is $0.21 \pm 0.05$.  Hence, we can consider the ejecta as an incompressible medium and treat its environment as a cool plasma ($\beta_{\rm e} = 0$).  We shall skip the derivation of wave dispersion relation (the reader can find its derivation in Zhelyazkov \& Chandra 2014) and will yield its final form
\begin{eqnarray}
\label{eq:dispeq}
	\frac{\left( \Omega^2 -
    \omega_{\rm Ai}^2 \right)F_m(\kappa_{\rm i}a) - 2mA \omega_{\rm Ai}/\sqrt{\mu \rho_{\rm i}}}
    {\left( \Omega^2 -
    \omega_{\rm Ai}^2 \right)^2 - 4A^2\omega_{\rm Ai}^2/\mu \rho_{\rm i} } \nonumber \\
    \nonumber \\
    {} = \frac{P_m(\kappa_{\rm e} a)}
    {{\displaystyle \frac{\rho_{\rm e}}{\rho_{\rm i}}} \left( \omega^2 - \omega_{\rm Ae}^2
    \right) + A^2  P_m(\kappa_{\rm e} a)/\mu \rho_{\rm i}},
\end{eqnarray}
where, $\Omega = \omega - \vec{k}\cdot \vec{v}_0$ is the Doppler-shifted wave frequency in the moving flux tube,
\[
	\kappa_{\rm i} = k_z\left[  1 - 4 A^2 \omega_{\rm Ai}^2/
        \mu \rho_{\rm i} \left( \Omega^2 -
        \omega_{\rm Ai}^2\right)^2 \right]^{1/2},\qquad \kappa_{\rm e} = k_z \left( 1 - \omega^2/\omega_{\rm Ae}^2 \right)^{1/2},
\]
\[
    \omega_{\rm Ai} = \frac{mA + k_z B_{{\rm i}z}}{\sqrt{\mu \rho_{\rm i}}}, \qquad \omega_{\rm Ae} = \frac{k_z B_{{\rm e}z}}{\sqrt{\mu \rho_{\rm e}}} = k_z v_{\rm Ae} \quad \mbox{with} \quad v_{\rm Ae} = B_{\rm e}/\sqrt{\mu \rho_{\rm e}},
\]
and
\[
    F_m(\kappa_{\rm i}a) = \frac{\kappa_{\rm i}a I_m^{\prime}(\kappa_{\rm i}a)}{I_m(\kappa_{\rm i}a)} \quad \mbox{and} \quad P_m(\kappa_{\rm e}a) = \frac{\kappa_{\rm e}a K_m^{\prime}(\kappa_{\rm e}a)}{K_m(\kappa_{\rm e}a)}.
\]
Note that $\omega_{\rm Ai}$ and $\omega_{\rm Ae}$ are the corresponding local Alfv\'en frequencies, and prime sign means a differentiation by the Bessel function argument.
Dispersion equation (\ref{eq:dispeq}) is similar to the dispersion equation of normal MHD modes in a twisted flux tube surrounded by incompressible plasma (Zhelyazkov \& Zaqarashvili 2012---there, in Eq.~(13), $\kappa_{\rm e} = k_z$), and to the dispersion equation for a twisted tube with non-magnetized environment, that is, with $\omega_{\rm Ae} = 0$ (Zaqarashvili \emph{et al.}\ 2010).

\subsection{Numerical results and comparison with observational data}
\label{subsec:numresults}
The main goal of our study is to determine under which conditions the MHD waves propagating along the moving flux tube can become unstable.  To conduct this investigation, it is necessary to assume that the wave frequency $\omega$ is a complex quantity, that is, $\omega \to \omega + \mathrm{i}\gamma$, where $\gamma$ is the instability growth rate, while the longitudinal wave number $k_z$ is a real variable in the wave dispersion relation.  Since the occurrence of the expected KH instability is determined primarily by the jet velocity, by searching for a critical or threshold value of it, we will gradually change its magnitude from zero to that critical value (and beyond).  Thus, we have to solve dispersion relations in complex variables, obtaining the real and imaginary parts of the wave frequency, or as is commonly accepted, of the wave phase velocity $v_{\rm ph} = \omega/k_z$, as functions of $k_z$ at various values of the velocity shear between the surge and its environment, $v_0$.

\begin{figure}[ht]
  \centering
\subfigure{\includegraphics[width = 2.95in]{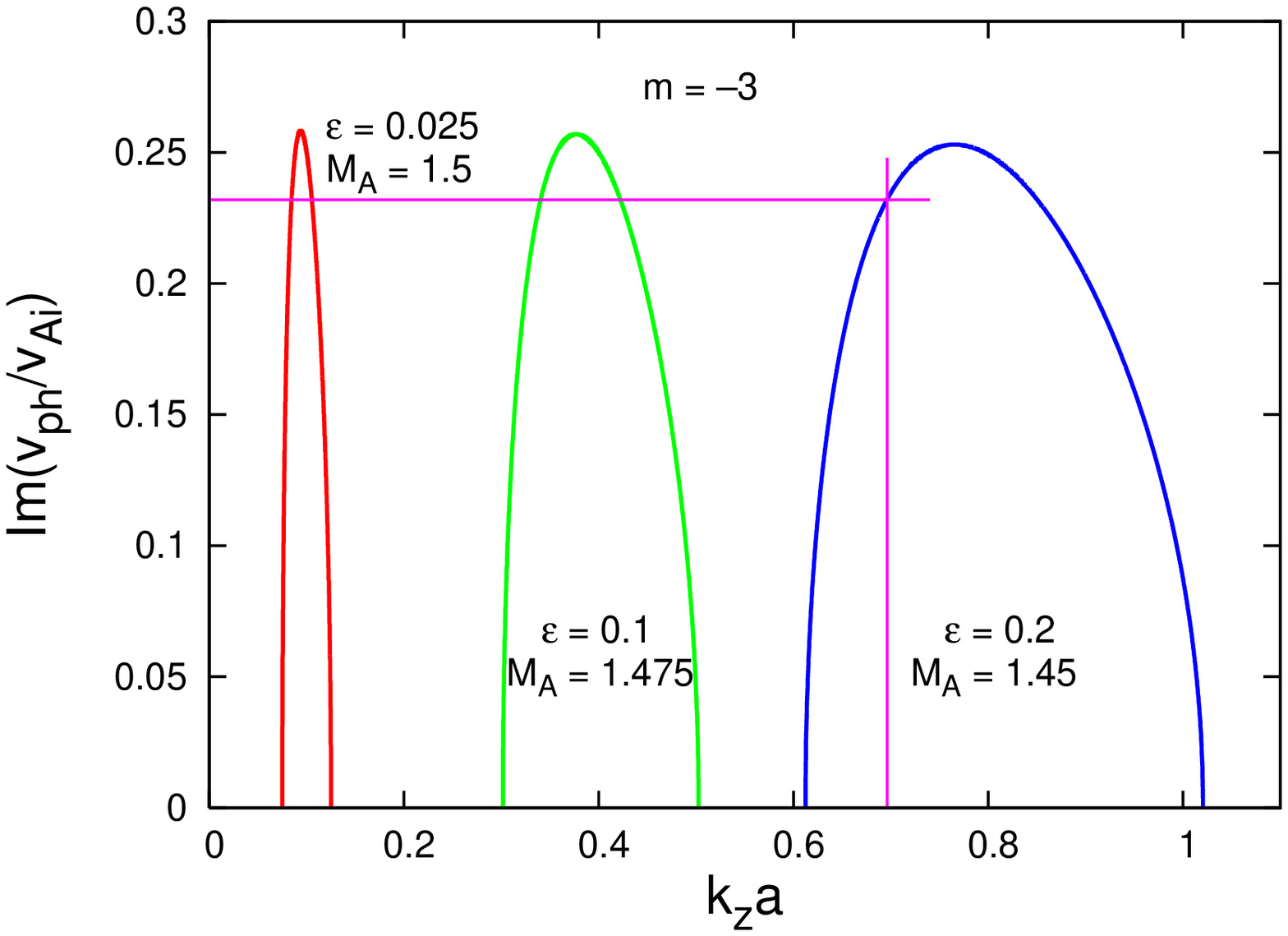}} \\
\subfigure{\includegraphics[width = 2.95in]{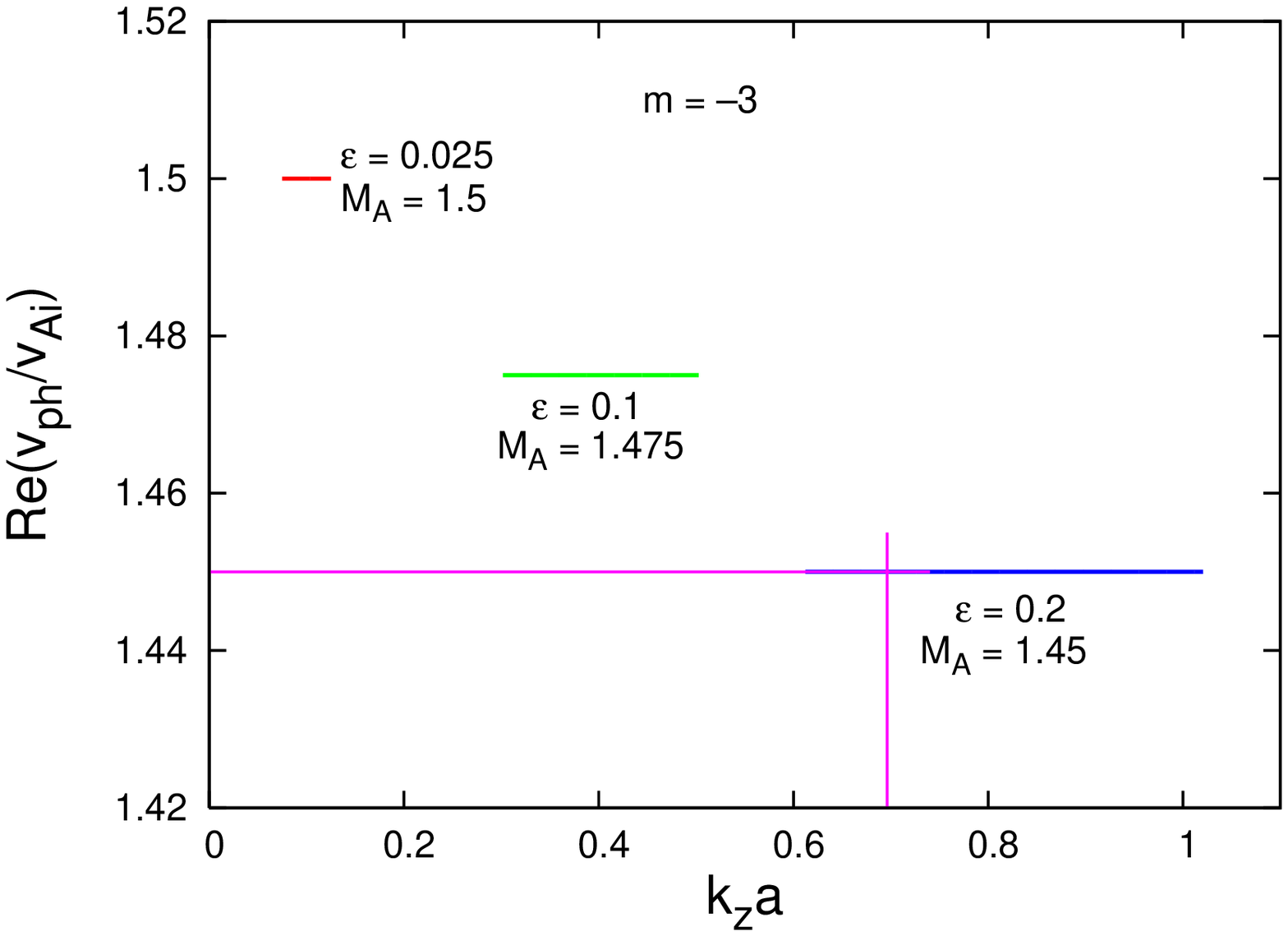}}
  \caption{(\emph{Top panel}) Growth rates of an unstable $m = -3$ MHD mode in three instability windows.  Adequate numerical values consistent with the observational data are obtained at $k_z a = 0.696245$, located in the third instability window, for which value of $k_z a$ the normalized mode growth rate is equal to $0.232117$. (\emph{Bottom panel}) Dispersion curves of unstable $m = -3$ MHD mode for three values of the twist parameter $\varepsilon = 0.025$, $0.1$, and $0.2$. The normalized phase velocity at $k_z a = 0.696245$ is equal to $1.45$.  (Adopted from Zhelyazkov, Zaqarashvili \& Chandra 2015.)}
  \label{fig:fig4}
\end{figure}
We shall solve Eq.~(\ref{eq:dispeq}) numerically and the input parameters of that task are the density contrast between the tube and its environment, $\eta = \rho_{\rm e}/\rho_{\rm i}$, and the twisted magnetic field by the ratio of the two magnetic field components, $\varepsilon = B_{{\rm i}\varphi}/B_{{\rm i}z}$, evaluated at the inner boundary of the tube, $r = a$, that is, with $B_{{\rm i}\varphi} = Aa$.  We normalize the velocities to the Alfv\'en speed $v_{\rm Ai} = B_{{\rm i}z}/(\mu \rho_{\rm i})^{1/2}$, thus, defining the dimensionless wave phase velocity $v_{\rm ph}/v_{\rm Ai}$ and the Alfv\'en Mach number $M_{\rm A} = v_0/v_{\rm Ai}$, the latter characterizing the axial motion of the tube.  The wavelength, $\lambda = 2\pi/k_z$, is normalized to the tube radius $a$ which implies that the dimensionless wave number is $k_z a$.  We note that the normalization of Alfv\'en frequency outside the jet, $\omega_{\rm Ae}$, requires except the tube radius, $a$, and the density contrast, $\eta$, the ratio of the two axial magnetic fields, $b = B_{\rm e}/B_{{\rm i}z}$.

Our choice for the density contrast is $\eta = 0.88$, which corresponds to electron densities $n_{\rm i} = 8.7 \times 10^8$~cm$^{-3}$ and $n_{\rm e} = 7.67 \times 10^8$~cm$^{-3}$.  With $\beta_{\rm i} = 1.5$ and $\beta_{\rm e} = 0$, the ratio of axial magnetic fields is $b = 1.58$.  If one fixes the Alfv\'en speed in the environment to be $v_{\rm Ae} \cong 787$~km\,s$^{-1}$ (that is, the value corresponding to $n_{\rm e} = 7.67 \times 10^8$~cm$^{-3}$ and $B_{\rm e} = 10$~G), the total pressure balance equation at $\eta = 0.88$ requires a sound speed inside the jet $c_{\rm si} \cong 523$~km\,s$^{-1}$ and Alfv\'en speed $v_{\rm Ai} \cong 467$~km\,s$^{-1}$ (more exactly, $467.44$~km\,s$^{-1}$), which corresponds to a magnetic field in the flux tube $B_{{\rm i}z} = 6.32$~G.  Following Ruderman (2007), to satisfy the Shafranov--Kruskal stability criterion for a kink instability we assume that the azimuthal component of the magnetic field $\vec{B}_{\rm i}$ is smaller than its axial component, meaning that we choose our twist parameter $\varepsilon$ to be always lower than $1$.  Computations show (see Fig.~1 and corresponding discussion in Zhelyazkov \& Chandra 2014) that the kink mode, $m = 1$, can become unstable against KH instability if the speed of moving tube exceeds ${\cong}1380$~km\,s$^{-1}$---a speed being in principal accessible for CMEs, but in fact, more than $2$ times higher than the registered by Foullon \emph{et al.}\ (2013) threshold speed of $680$~km\,s$^{-1}$.
Hence, detected KH instability cannot be associated with the kink mode.  But the situation distinctly changes for the $m = -3$ MHD mode (Zhelyazkov, Zaqarashvili \& Chandra 2015).  Figure~\ref{fig:fig4} shows the appearance of three instability windows on the $k_z a$-axis.  The width of each instability window depends upon the value of the twist parameter $\varepsilon$---the narrowest window corresponds to $\varepsilon = 0.025$, and the widest to $\varepsilon = 0.2$.  Note that the phase velocities of unstable $m = -3$ MHD modes coincide with the magnetic flux tube speeds (the bottom panel of Fig.~\ref{fig:fig4} shows that the normalized wave velocity on given dispersion curve is equal to its label $\mathsf{M}_{\sf A}$).  Therefore, unstable perturbations are frozen in the flow, and consequently, they are vortices rather than waves.  This is firmly based on physics because the KH instability in hydrodynamics deals with unstable vortices.  All critical Alfv\'en Mach numbers yield acceptable threshold speeds of the ejecta that ensure the occurrence of KH instability---these speeds are equal to $701$~km\,s$^{-1}$, $689$~km\,s$^{-1}$, and $678$~km\,s$^{-1}$ and agree very well with the speed of $680$~km\,s$^{-1}$ found by Foullon \emph{et al.}\ (2013).  The observationally detected KH instability wavelength $\lambda_{\rm KH} = 18.5$~Mm and ejecta width $\Delta L = 4.1$~Mm define the corresponding instability dimensionless wave number, $k_z a = \pi \Delta L/\lambda$, to be equal to $0.696245$.  Figure~\ref{fig:fig4} shows that $k_z a = 0.696245$ lies in the third instability window and accordingly determines a value of the dimensionless growth rate Im$(v_{\rm ph}/v_{\rm Ai}) = 0.232117$ (see the top panel of Fig.~\ref{fig:fig4}), which implies a computed wave growth rate $\gamma_{\rm KH} = 0.037$~s$^{-1}$, being in good agreement with the deduced from observations $\gamma_{\rm KH} = 0.033$~s$^{-1}$.  We also note that the wave phase velocity estimated from Fig.~\ref{fig:fig4} (bottom panel) of $678$~km\,s$^{-1}$ is rather close to the speed of the $131$~\AA~CME leading edge, which is equal to $687$~km\,s$^{-1}$.  The position of a given instability window, at fixed input parameters $\eta$ and $b$, is determined chiefly by the magnetic field twist in the moving flux tube.  This circumstance allows us by slightly shifting the third instability window to the right, to tune the vertical purple line (see the top panel of Fig.~\ref{fig:fig4}) to cross the growth rate curve at a value that would yield $\gamma_{\rm KH} = 0.033$~s$^{-1}$.  The necessary shift of only $0.023625$ can be achieved by taking the magnetic field twist parameter, $\varepsilon$, to be equal to $0.20739$---in that case the normalized Im$(v_{\rm ph}/v_{\rm Ai}) = 0.207999$ gives the registered KH instability growth rate of $0.033$~s$^{-1}$.  (That very small instability window shift does not change noticeably the critical ejecta speed.)  In this way, one demonstrates the flexibility of this model, which allows deriving the numerical KH instability characteristics in very good agreement with observational data.
\begin{figure}[ht]
  \centering
\subfigure{\includegraphics[width = 2.95in]{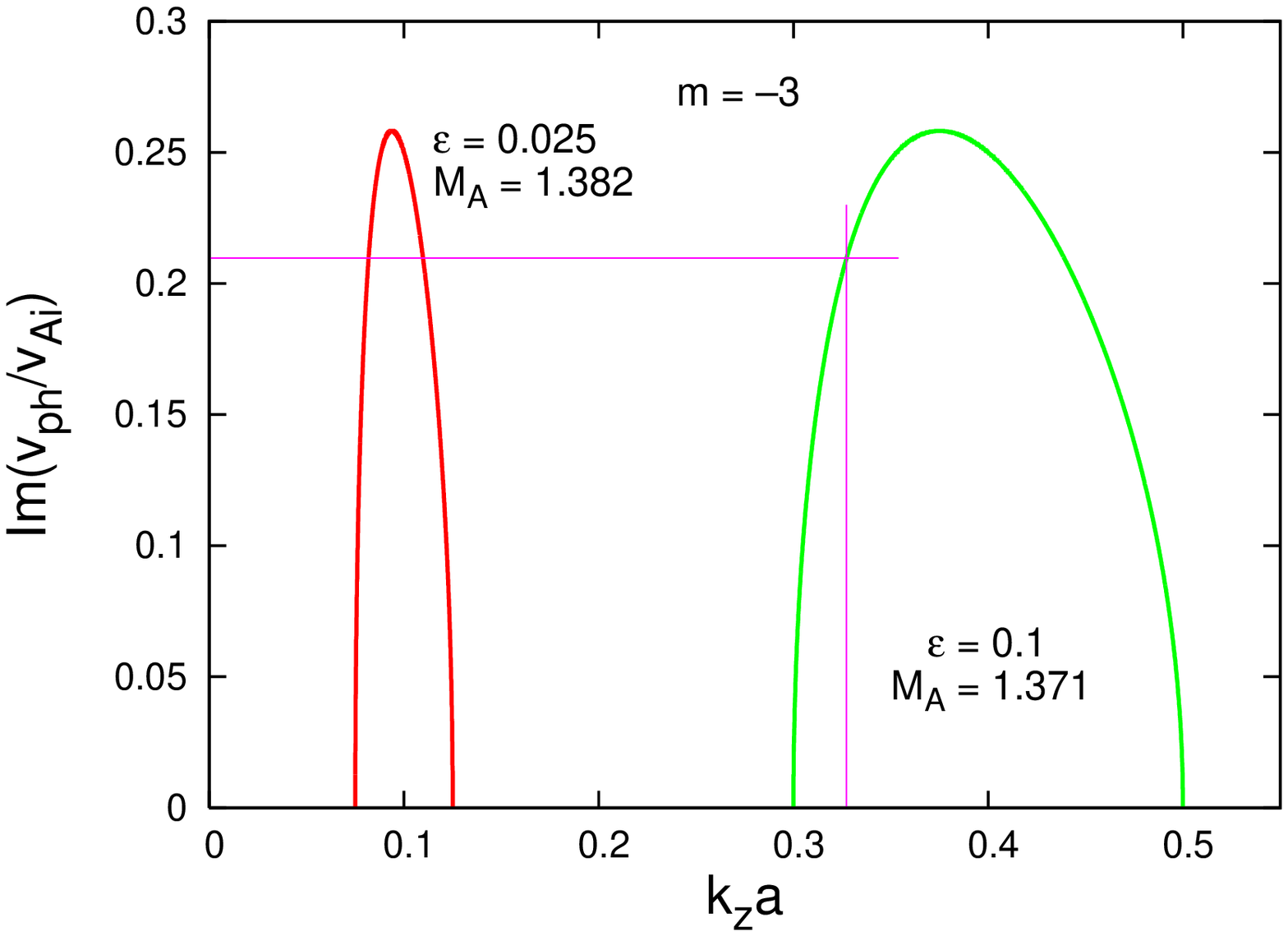}} \\
\subfigure{\includegraphics[width = 2.95in]{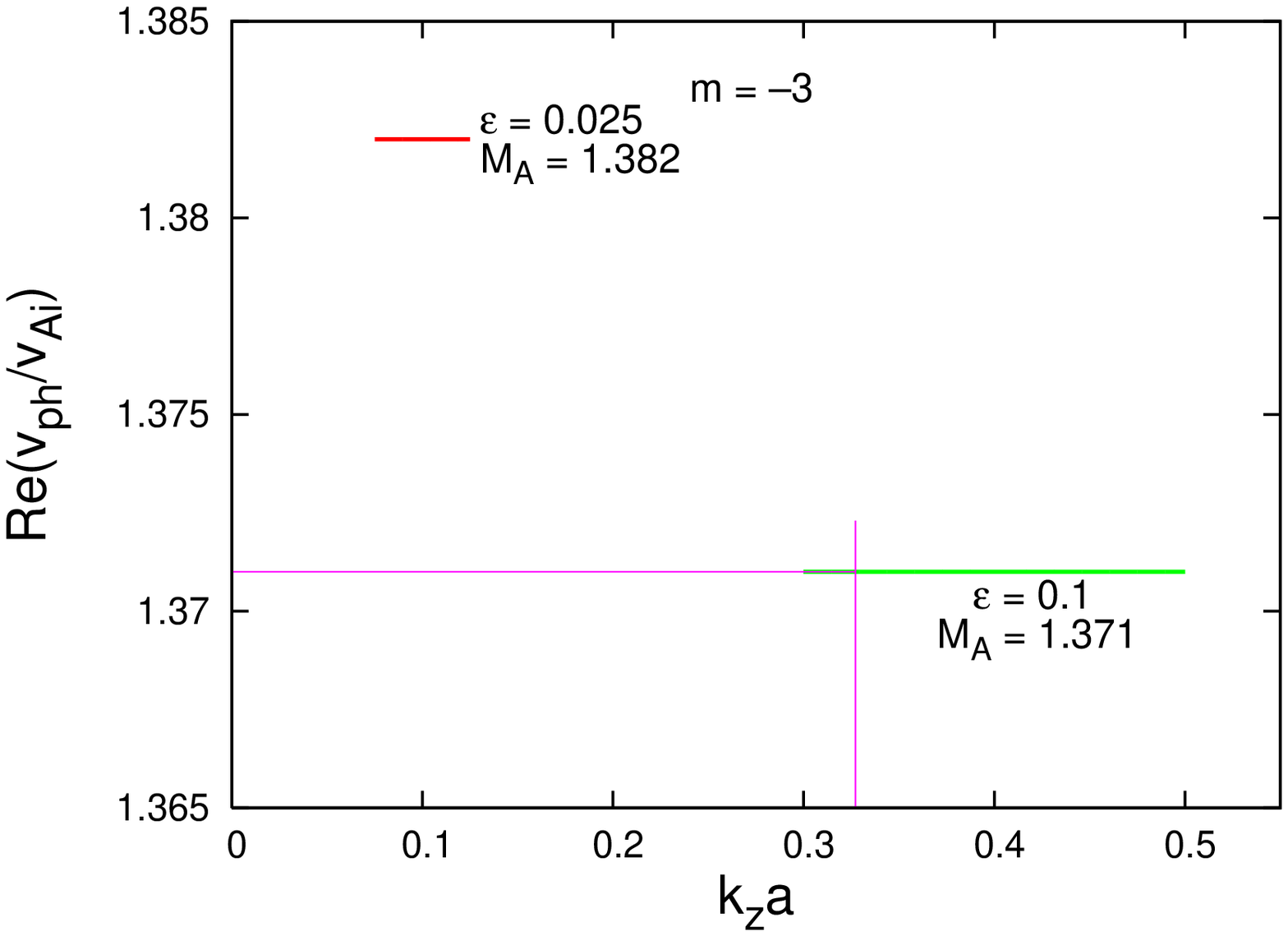}}
  \caption{(\emph{Top panel}) Growth rates of an unstable $m = -3$ MHD mode in two instability windows.  The best coincidence with the observational data one obtains at $k_z a = 0.32725$, for which value the normalized mode growth rate is equal to $0.2096$. (\emph{Bottom panel}) Dispersion curves of unstable $m = -3$ MHD mode for two values of the twist parameter $\varepsilon = 0.025$ and $0.1$. The normalized phase velocity at $k_z a = 0.32725$ is equal to $1.371$.}
  \label{fig:fig5}
\end{figure}

The observed quasi-periodic vortex-like structures at the northern boundary of a filament during an eruption on 2011 February 24 by M\"{o}stl, Temmer \& Veronig (2013) can be modeled in a similar way.  In their case the environment electron density $n_{\rm e}$ is taken to be $1 \times 10^9$~cm$^{-3}$, plasma pressure $p_{\rm e}$ is $0.09$~Pa, and the magnetic field $B_{\rm e}$ is $10$~G.  In the filament, they assumed that the density is at least $10$ times higher than the corona value (\emph{i.e.}, $n_{\rm i} = 1 \times 10^{10}$~cm$^{-3}$) and that $T_{\rm i} \leqslant 1 \times 10^5$~K.  From these data one obtains that the temperature of coronal plasma is $T_{\rm e} = 3.26 \times 10^6$~K and, accordingly, $c_{\rm se} \cong 212$~km\,s$^{-1}$.  Alfv\'en speed in the environment is $v_{\rm Ae} \cong 689$~km\,s$^{-1}$.  With a density contrast $\eta = 0.1$, the total pressure (sum of thermal and magnetic pressures) balance equation yields the Alfv\'en velocity inside the ejecta, $v_{\rm Ai} = 226.4$~km\,s$^{-1}$.  The magnetic field there is similar to that in the corona, $B_{{\rm i}z} \cong 10.4$~G, that implies a value of parameter $b = B_{\rm e}/B_{{\rm i}z}$ close to $1$ (actually it is equal to $1.0385$).  Considering both media as incompressible magnetized plasmas we can use dispersion equation (\ref{eq:dispeq}) with taking the wave attenuation coefficient in the environment, $\kappa_{\rm e}$, to be simply equal to $k_z$, that is, to use Eq.~(13) of Zhelyazkov \& Zaqarashvili (2012).  As in the case of Foullon \emph{et al.}\ 2013, the kink mode ($m = 1$) may become unstable at rather high speed of the moving flux tube, notably ${\cong}1138$~km\,s$^{-1}$---a speed that is ${\sim}3.7$ times higher than the detected critical velocity of $310$~km\,s$^{-1}$.  The $m = -3$ MHD harmonic (see the top panel in Fig.~\ref{fig:fig5}) becomes unstable at threshold Alfv\'en Mach numbers equal to $1.382$ (at $\varepsilon = 0.025$) and $1.371$ (at $\varepsilon = 0.1$).  According to M\"{o}stl, Temmer \& Veronig 2013, the half-width of the filament, $\Delta L/2 = a$, lies between $0.5$~Mm and $1.0$~Mm.  If one takes $a = 0.75$~Mm, the registered KH wavelength of $14.4$~Mm corresponds to the dimensionless wavenumber $k_z a \cong 0.327$.  Figure~\ref{fig:fig5} shows that this normalized wavenumber falls in the second instability window for a twist of $0.1$.  For $a = 1.0$~Mm, the corresponding $k_z a = 0.436$, which also alights in the same window.  However, for the smallest $a = 0.5$~Mm, the dimensionless wavenumber is equal to $0.218$ and one needs to perform calculations for a smaller value of the magnetic field twist parameter $\varepsilon$---a value of $0.08$ would yield an instability window accommodating that $k_z a = 0.218$.  The normalized wave phase velocity growth rate that corresponds to $k_z a \cong 0.327$ is equal (see the top panel in Fig.~\ref{fig:fig5}) to $0.2096$ which yields a linear growth rate $\gamma_{\rm KH} \cong 0.021$~s$^{-1}$.  For the other two dimensionless wavenumbers the corresponding growth rates are of the same order.  The computed critical filament speed in the second instability window is equal to $310.4$~km\,s$^{-1}$, that is, very close to the deduced from observations velocity of $310$~km\,s$^{-1}$.  The value of the wave phase velocity, as in the case of Foullon \emph{et al.}\ 2013, coincides with the critical speed of corresponding instability window.

\section{Conclusion and outlook}
\label{sec:concl}
A successful modeling of the KH instability in solar eruptive events like CMEs, spicules, surges, UV, EUV, and X-ray jets requires an adequate physical model.  The model used here in studying KH instability in two well documented CMEs is a cylindrical twisted magnetic flux tube that moves with a velocity of $\vec{v}_0$ with regard to the surrounding solar magnetized plasma.  Such a twisted flux tube is, however, a consequence of the evolution of a pre-launch magnetic flux rope.  Flux ropes form not only before (and sometimes after) CMEs, but also in the solar wind (Cid \emph{et al.}\ 2001; Zaqarashvili \emph{et al.}\ 2014b), in the Earth's magnetotail (Hietala, Eastwood \& Isavnin 2014), and solar tornados (Zhang \& Liu 2011; Li \emph{et al.}\ 2012; Wedemeyer-B\"{o}hm \emph{et al.}\ 2012).  We also note that a flux rope eruption can be triggered by recurrent chromospheric plasma injections (surges/jets) as was reported by Guo \emph{et al.}\ (2010).  Their study confirms that the surge activities can efficiently supply the necessary material for some filament formation.  Furthermore, the study indicates that the continuous mass with momentum loaded by the surge activities to the filament channel could make the filament unstable and cause it to erupt.  In all these cases when a flux rope is formed we can believe that a twisted magnetic flux structure is the final form of the solar eruption.

The KH vortices imaged by Foullon \emph{et al.}\ (2013) on the 2010 November 3 C-type CME can be explained as a KH instability of the $m = -3$ harmonic in a twisted flux tube moving in external cool magnetized plasma embedded in homogeneous untwisted magnetic field.  We have assumed the wave vector $\vec{k}$ to be aligned with the $\vec{v}_0 = \vec{V}_{\rm i} - \vec{V}_{\rm e}$ vector.  We would like to point out that the results of the numerical modeling crucially depend on the input parameters.  Any small change in the density contrast, $\eta$, or the magnetic fields ratio, $b$, can dramatically change the picture.  In considered case, the input parameters for solving the MHD mode dispersion relation in complex variables (the mode frequency $\omega$ and, respectively, the mode phase velocity $v_{\rm ph} = \omega/k_z$, were considered as complex quantities) were chosen to be consistent with the plasma and magnetic field parameters listed in Table~2 in Foullon \emph{et al.}\ 2013.    With a twist parameter of the background magnetic filed $\vec{B}_{\rm i}$, $\varepsilon = 0.2$, the critical jet's speed is $v_0^{\rm cr} = 678$~km\,s$^{-1}$, and at a wavelength of unstable $m = -3$ mode $\lambda_{\rm KH} = 18.5$~Mm and ejecta width $\Delta L = 4.1$~Mm, its growth rate is $\gamma_{\rm KH} = 0.037$~s$^{-1}$.  These values of $v_0^{\rm cr}$ and $\gamma_{\rm KH}$ agree well with the data listed in Table~3 of Foullon \emph{et al.}\ 2013.  We also showed that the numerically obtained instability growth rate can be slightly reduced to coincide with the observational rate of $0.033$~s$^{-1}$ through slightly shifting the appropriate instability window to the right---this can be done by performing the calculations with a new value of the magnetic field twist parameter $\varepsilon$, equal to $0.20739$.  Thus, this model is flexible enough to allow us to numerically derive KH instability characteristics very close to the observed ones.  Critical ejecta speed and KH instability growth rate values, in good agreement with those derived by Foullon \emph{et al.}\ (2013), can also be obtained by exploring the fluting-like, $m = -2$, MHD mode (Zhelyazkov \& Chandra 2014).  Although the KH instability characteristics of both $m = -2$ and $m = -3$ MHD modes agree well with observational data, only the instability of the $m = -3$ harmonic may explain why the KH vortices are seen only at one side of rising CME (Foullon \emph{et al.}\ 2013).  This harmonic yields that the unstable vortices have three maxima around the magnetic tube with a $360/3 = 120$ degree interval.  Therefore, if one maximum is located in the plane perpendicular to the line of sight (as it is clearly seen by observations), then one cannot detect two other maxima in imaging observations because they will be largely directed along the line of sight.

The modeling S-type coronal mass ejection on 2011 February 24 detected by M\"{o}stl, Temmer \& Veronig (2013) carried out in a similar way yields a critical speed of $310.4$~km\,s$^{-1}$, practically coinciding with the observed one.  Furthermore, the model predicts that the linear growth rate of the unstable $m = -3$ MHD harmonic with a wavelength $\lambda_{\rm KH} = 14.4$~Mm is $\gamma_{\rm KH} = 0.021$~s$^{-1}$---it remains this value to be validated.  Although similar in nature, the two ejecta (Foullon \emph{et al.}\ 2013; M\"{o}stl, Temmer \& Veronig 2013) have distinctive characteristics: the density contrast in Foullon \emph{et al.}\ 2013 is small ($\eta = 0.88$) while in M\"{o}stl, Temmer \& Veronig 2013 it is relatively high ($\eta = 0.1$).  The first ejecta is much hotter that its environment, while for the second one we have just the opposite situation.  It is necessary to stress that each CME is a unique event and its successful modeling requires a full set of observational data for the plasma densities, magnetic fields, and temperatures of both media along with the detected ejecta width and speed.  Concerning values of the flux tube speeds at which the instability starts, they can vary from a few kilometers per second, $6$--$10$~km\,s$^{-1}$, as observed by Ofman \& Thompson (2011), through $310 \pm 20$~km\,s$^{-1}$ of M\"{o}stl, Temmer \& Veronig (2013), to $680 \pm 92$~km\,s$^{-1}$ deduced from Foullon \emph{et al.}\ (2013).  It is curious to see whether the $13$ fast flareless CMEs (with velocities of $1000$~km\,s$^{-1}$ and higher) observed from 1998 January 3 to 2005 January 4 (see Table~1 in Song \emph{et al.}\ 2013), are subject to the KH instability.

This approach---exploring the conditions for emerging a KH instability of MHD high-modes in solar atmosphere eruptions---has to be applied to the recent observations of oscillations and waves in spicules and macro-spicules (\emph{e.g.}, Popescu \emph{et al.}\ 2007; Zaqarashvili 2011; Tavabi, Koutchmy, \& Ajabshirizadeh 2011; Tsiropoula \emph{et al.}\ 2012; Pereira, De Pontieu \& Carlsson 2013; Skogsrud, Rouppe van der Voort \& De Pontieu 2014), mottles (Kuridze \emph{et al.}\ 2012, 2013), chromospheric evaporations and jets (Kuridze \emph{et al.}\ 2011;  Doschek, Warren \& Young 2013; Yurchyshyn \emph{et al.}\ 2014), surges (Liu 2008; Zheng \emph{et al.}\ 2013), and coronal X-ray jets (Chandrashekhar \emph{et al.}\ 2014a, 2014b).  No less challenging is the modeling of the KH instability in solar prominences with rotating motions (Su \& van Ballegooijen 2013) and in solar tornado-like prominences (Panasenco, Martin \& Velli 2014).  The first step in this direction is already done---very recently, Zhelyazkov \emph{et al.}\ (2014) modeled the KH instability in a high-temperature surge observed by Kayshap, Srivastava \& Murawski (2013) and showed that the $m = -2$ and $m = -3$ MHD modes can become unstable when the magnetic field twist parameter has values between $0.2$ and $0.4$.  The critical jet velocities for instability onset lie in the range of $93.5$--$99.3$~km\,s$^{-1}$, and the instability growth rate, depending on the value of the wavelength, is of the order of several dozen inverse milliseconds.

\vspace{5mm}

\textbf{Acknowledgments} \hspace{2mm}
This work was supported by the Bulgarian Science Fund and the Department of Science \& Technology, Government of India Fund under Indo-Bulgarian bilateral project CSTC/INDIA 01/7, /Int/ Bul\-garia/P-2/12.  The author is indebted to the referee for his constructive criticism and valuable suggestions for clarifying the paper presentation, Teimuraz Zaqarashvili, Abhishek Srivastava, and Ramesh Chandra for fruitful discussions, and Snezhana Yordanova for drawing one figure.

\vspace{5mm}

\centerline{\large \textbf{References}}

\vspace{2mm}

\noindent Andries, J., Goossens, M.\ 2001, \emph{A\&A\/} \textbf{368}, 1083--1094.

\noindent Antolin, P., Shibata, K., Kudoh, T., Shiota, D., Brooks, D.\ 2009, in: \emph{ASP \\
\hspace*{3mm} Conference Series\/} \textbf{415}, B.~Lites, M.~Cheung, T.~Magara, J.~Mariska, and \\
\hspace*{3mm} K.~Reeves (eds), ASP, San Francisco, California, pp.~247--251.

\noindent Aschwanden, M.~J., Fletcher, L., Schrijver, C.~J., Alexander, D.\ 1999,\\
\hspace*{3mm} \emph{Astrophys.\ J.}\ \textbf{520}, 880--894.

\noindent Aschwanden, M.~J., Schrijver, C.~J.\ 2011, \emph{Astrophys.\ J.}\ \textbf{736}, 102 (20pp).

\noindent Aulanier, G.\ 2014, in: \emph{Proc. IAU Symposium No.~300}, B.~Schmieder, J.-M. \\
\hspace*{3mm} Malherbe, and S.~T.~Wu (eds), Cambridge University Press, New York,\\
\hspace*{3mm} pp.~184--196.

\noindent Aulanier, G., T\"{o}r\"{o}k, T., D\'emoulin, P., DeLuca, E.~E.\ 2010, \emph{Astrophys.\ J.}\ \\
\hspace*{3mm} \textbf{708}, 314--333.

\noindent Beckers, J.~M.\ 1972, \emph{Ann.\ Rev.\ Astron.\ Astrophys.}\ \textbf{10}, 73--100.

\noindent Bennett, K., Roberts, B., Narain, U.\ 1999, \emph{Solar Phys.}\ \textbf{185}, 41--59.

\noindent Bogdan, T.~J.\ 1984, \emph{Astrophys.\ J.}\ \textbf{282}, 769--775.

\noindent Bohlin, J.~D., Vogel, S.~N., Purcell, J.~D., Sheeley, Jr., N.~R., Tousey, R.,\\
\hspace*{3mm} VanHoosier, M.~E.\ 1975, \emph{Astrophys.\ J.}\ \textbf{197}, L133--L135.

\noindent Brueckner, G.~E., Bartoe, J.-D.~F.\ 1983, \emph{Astrophys.\ J.}\ \textbf{272}, 329--348.

\noindent Brueckner, G.~E., Howard, R.~A., Koomen, M.~J.\ 1995, \emph{Solar Phys.}\ \textbf{162},\\
\hspace*{3mm} 357--402.

\noindent Budnik, F., Schr\"{o}der, K.-P., Wilhelm, K., Glassmeier, K.-H.\ 1998, \emph{A\&A\/} \\
\hspace*{3mm} \textbf{334}, L77--L80.

\noindent Carter, B.~K., Erd\'elyi, R.\ 2008, \emph{A\&A\/} \textbf{481}, 239--246.

\noindent Cavus, H., Kazkapan, D.\ 2013, \emph{New Astron.}\ \textbf{25}, 85--94.

\noindent Chandra, R., Gupta, G.~R., Mulay, S., Tripathi, D.\ 2014, \emph{MNRAS\/} \textbf{446},\\
\hspace*{3mm} 3741--3748.

\noindent Chandra, R., Schmieder, B., Mandrini, C.~H., D\'emoulin, P., Pariat E.,\\
\hspace*{3mm} T\"{o}r\"{o}k, T., Uddin, W.\ 2011, \emph{Solar Phys.}\ \textbf{269}, 83--104.

\noindent Chandrashekhar, K., Bemporad, A., Banerjee, D., Gupta, G.~R., Teriaca, L.\ \\
\hspace*{3mm} 2014a, \emph{A\&A\/} \textbf{561}, A104 (9pp).

\noindent Chandrashekhar, K., Morton, R.~J., Banerjee, D., Gupta, G.~R.\ 2014b, \emph{A\&A\/}\\
\hspace*{3mm} \textbf{562}, A98 (10pp).

\noindent Chen, B., Bastian, T.~S., Gary, D. E.\ 2014, \emph{Astrophys.\ J.}\ \textbf{794}, 149 (14pp).

\noindent Cheng, X., Ding, M.~D., Guo, Y., Zhang, J., Vourlidas, A., Liu, Y.~D.,\\
\hspace*{3mm} Olmedo, O., Sun, J.~Q., Li, C.\ 2014, \emph{Astrophys.\ J.}\ \textbf{780}, 28 (9pp).

\noindent Cheng, X., Zhang, J., Liu, Y., Ding, M.~D.\ 2011, \emph{Astrophys.\ J.}\ \textbf{732}, L25 \\
\hspace*{3mm} (6pp).

\noindent Cid, C., Hidalgo, M.~A., Sequeiros, J., Rodriguez-pacheco, J., Bronchalo, E.\ \\
\hspace*{3mm} 2001, \emph{Solar Phys.}\ \textbf{198}, 169--177.

\noindent Cirtain, J.~W., Golub,~L., Lundquist, L., \emph{et al.}\ 2007, \emph{Science\/} \textbf{318}, \\
\hspace*{3mm} 1580--1582.

\noindent Defouw, R.~J.\ 1976, \emph{Astrophys.\ J.}\ \textbf{209}, 266--269.

\noindent De Pontieu, B., Hansteen, V.~H., Rouppe van der Voort, L., van Noort, M.,\\
\hspace*{3mm} Carlsson, M.\ 2007a, in \emph{ASP Conference Series\/} \textbf{368}, P.~Heinzel, I.~Doroto- \\
\hspace*{3mm} vi\v{c}, and R.~J.~Rutten (eds), ASP, San Francisco, California, pp.~65--80.

\noindent De Pontieu, B., McIntosh, S.~W., Carlsson, M., \emph{et al.}\ 2007b, \emph{Science\/} \\
\hspace*{3mm} \textbf{318}, 1574--1577.

\noindent De Pontieu, B., McIntosh, S., Hansteen, V.~H., \emph{et al.}\ 2007c, \emph{Publ.\ Astron.\ \\
\hspace*{3mm} Soc.\ Japan\/} \textbf{59}, S655--S662.

\noindent De Pontieu, B., Carlsson, M., Rouppe van der Voort, L.~H.~M., Rutten, \\
\hspace*{3mm} R.~J., Hansteen, V.~H., Watanabe, H.\ 2012, \emph{Astrophys.\ J.}\ \textbf{752}, L12 \\
\hspace*{3mm} (6pp).

\noindent D\'{i}az, A.~J., Oliver, R., Ballester, J.~L., Soler, R.\ 2011, \emph{A\&A\/} \textbf{533}, A95 \\
\hspace*{3mm} (12pp).

\noindent Doschek, G.~A., Warren, H.~P., Young, P.~R.\ 2013, \emph{Astrophys.\ J.}\ \textbf{767}, 55 \\
\hspace*{3mm} (13pp).

\noindent Dungey, J.~W., Loughhead, R.~E.\ 1954, \emph{Aust.\ J.\ Phys.}\ \textbf{7}, 5--13.

\noindent Edwin, P.~M., Roberts, B.\ 1983, \emph{Solar Phys.}\ \textbf{88}, 179--191.

\noindent Erd\'elyi, R., Carter, B.~K.\ 2006, \emph{A\&A\/} \textbf{455}, 361--370.

\noindent Erd\'elyi, R., Fedun, V.\ 2006, \emph{Solar Phys.}\ \textbf{238}, 41--59.

\noindent Erd\'elyi, R., Fedun, V.\ 2007, \emph{Solar Phys.}\ \textbf{246}, 101--116.

\noindent Erd\'elyi, R., Fedun, V.\ 2010, \emph{Solar Phys.}\ \textbf{263}, 63--85.

\noindent Evans, R.~M., Pulkkinen, A.~A., Zheng, Y., Mays, M.~L., Taktakishvili, A.,\\
\hspace*{3mm} Kuznetsova, M.~M., Hesse, M.\ 2013, \emph{Space Weather\/} \textbf{11}, 333--334.

\noindent Forbes, T.~G., Linker, J.~A., Chen, J., \emph{et al.}\ 2006, \emph{Space Sci.\ Rev.}\ \textbf{123},\\
\hspace*{3mm} 251--302.

\noindent Foullon, C., Verwichte, E., Nakariakov, V.~M., Nykyri, K., Farrugia, C.~J.\ \\
\hspace*{3mm} 2011, \emph{Astrophys.\ J.}\ \textbf{729}, L8 (4pp).

\noindent Foullon, C., Verwichte, E., Nykyri, K., Aschwanden, M.~J., Hannah, I.~G.\ \\
\hspace*{3mm} 2013, \emph{Astrophys.\ J.}\ \textbf{767}, 170 (18pp).

\noindent Furno, I., Intrator, T.~P., Lapenta, G., Dorf, L., Abbate, S., Ryutov, D.~D.\ \\
\hspace*{3mm} 2007, \emph{Phys.\ Plasmas\/} \textbf{14}, 022103 (10pp).

\noindent Goossens, M., Hollweg, J.~V., Sakurai, T.\ 1992, \emph{Solar Phys.}\ \textbf{138}, 233--255.

\noindent Goossens, M., Terradas, J., Andries, J., Arregui, I., Ballester, J.~L.\ 2009, \\
\hspace*{3mm} \emph{A\&A\/} \textbf{503}, 213--223.

\noindent Goossens, M., Andries, J., Soler, R., Van Doorsselaere, T., Arregui, I.,\\
\hspace*{3mm} Terradas, J.\ 2012, \emph{Astrophys.\ J.}\ \textbf{753}, 111 (12pp).

\noindent Gosling, J.~T., Hildner, E., MacQueen, R.~M., Munro, R.~H., Poland, A.~I.,\\
\hspace*{3mm} Ross, C.~L.\ 1974, \emph{J.\ Geophys.\ Res.: Space Phys.}\ \textbf{79}, 4581--4587.

\noindent Green, L.~M., Kliem, B.\ 2014, in: \emph{Proc. IAU Symposium No.~300}, B.~Schmieder,\\
\hspace*{3mm} J.-M.~Malherbe, and S.~T.~Wu (eds), Cambridge University Press, New \\
\hspace*{3mm} York, pp.~209--214.

\noindent Guo, J., Liu, Y., Zhang, H., Deng, Y., Lin, J., Su, J.\ 2010, \emph{Astrophys.\ J.}\ \\
\hspace*{3mm} \textbf{711}, 1057--1061.

\noindent Guo, Y.,  Erd\'elyi, R., Srivastava, A.~K., Hao, Q., Cheng, X., Chen, P.~F.,\\
\hspace*{3mm} Ding, M.~D., Dwivedi, B.~N.\ 2015, \emph{Astrophys.\ J.}\ \textbf{799}, 151 (10pp).

\noindent Handy, B.~N., Acton, L.~W., Kankelborg, C.~C., \emph{et al.}\ 1999, \emph{Solar Phys.}\ \\
\hspace*{3mm} \textbf{187}, 229--260.

\noindent Hansteen, V.~H., De Pontieu, B., Rouppe van der Voort, L., van Noort, M.,\\
\hspace*{3mm} Carlsson, M.\ 2006, \emph{Astrophys.\ J.}\ \textbf{647}, L73--L76.

\noindent Hietala, H., Eastwood, J.~P., Isavnin, A.\ 2014, \emph{Plasma Phys.\ Control.\ Fusion\/} \\
\hspace*{3mm} \textbf{56}, 064011 (9pp).

\noindent Hood, A.~W., Priest, E.~R.\ 1979, \emph{Solar Phys.}\ \textbf{64}, 303--321.

\noindent Hollweg, J.~V.\ 1985, in: \emph{Advances in Space Plasma Physics}, B.~Buti (ed),\\
\hspace*{3mm} World Scientific, Singapore, pp.~77--141.

\noindent Howard, T.~A., DeForest, C.~E.\ 2014, \emph{Astrophys.\ J.}\ \textbf{796}, 33 (15pp).

\noindent Howard, T.~A., Webb, D.~F., Tappin, S.~J., Mizuno, D.~R., Johnston, J.~C.\ \\
\hspace*{3mm} 2006, \emph{J.\ Geophys.\ Res.: Space Phys.}\ \textbf{111}, A04105 (8pp).

\noindent Howard, T.~A., Moses, J.~D., Vourlidas, A., \emph{et al.}\ 2008, \emph{Space Sci.\ Rev.}\ \\
\hspace*{3mm} \textbf{136}, 67--115.

\noindent Jel\'{i}nek, P., Srivastava, A.~K., Murawski, K., Kayshap, P., Dwivedi, B.~N.\ \\
\hspace*{3mm} 2015, \emph{A\&A\/}, submitted.

\noindent Joshi, N.~C., Srivastava, A.~K., Filippov, B., Kayshap, P., Uddin, W., \\
\hspace*{3mm} Chandra, R.\ 2013a, \emph{Astrophys.\ J.}\ \textbf{771}, 65 (13pp).

\noindent Joshi, N.~C., Uddin, W., Srivastava, A.~K., \emph{et al.}\  2013b, \emph{Adv.\ Space Res.}\ \textbf{52},\\
\hspace*{3mm} 1--14.

\noindent Kamio, S., Curdt, W., Teriaca, L., Inhester, B., Solanki, S.~K.\ 2010, \emph{A\&A\/}\\
\hspace*{3mm} \textbf{510}, L1 (4pp).

\noindent Kayshap, P., Srivastava, A.~K., Murawski, K.\ 2013, \emph{Astrophys.\ J.}\ \textbf{763}, 24 \\
\hspace*{3mm} (12pp).

\noindent Kayshap, P., Srivastava, A.~K., Murawski, K., Tripathi, D.\ 2013, \emph{Astrophys.} \\
\hspace*{3mm} \emph{J.}\ \textbf{770}, L3 (8pp).

\noindent Kukhianidze, V., Zaqarashvili, T.~V., Khutsishvili, E.\ 2006, \emph{A\&A\/} \textbf{449},\\
\hspace*{3mm} L35--L38.

\noindent Kudoh, T., Shibata, K.\ 1999, \emph{Astrophys.\ J.}\ \textbf{514}, 493--505.

\noindent Kuridze, D., Mathioudakis, M., Jess, D.~B., Shelyag, S., Christian, D.~J., \\
\hspace*{3mm} Keenan, F.~P., Balasubramaniam, K.~S.\ 2011, \emph{A\&A\/} \textbf{533}, A76 (5pp).

\noindent Kuridze, D., Morton, R.~J., Erd\'elyi, R., Dorrian, G.~D., Mathioudakis,\\
\hspace*{3mm} M., Jess, D.~B., Keenan, F.~P.\ 2012, \emph{Astrophys.\ J.}\ \textbf{750}, 51 (5pp).

\noindent Kuridze, D., Verth, G., Mathioudakis, M., Erd\'elyi, R., Jess, D.~B., Morton,\\
\hspace*{3mm} R.~J., Christian, D.~J.,  Keenan, F.~P.\ 2013, \emph{Astrophys.\ J.}\ \textbf{799}, 82 (8pp).

\noindent Landi, E., Miralles, M.~P.\ 2014, \emph{Astrophys.\ J.}\ \textbf{780}, L7 (5pp).

\noindent Lemen, J.~R., Title, A.~M., Akin, D.~J., \emph{et al.}\ 2012, \emph{Solar Phys.}\ \textbf{275}, 17--40.

\noindent Li, X., Morgan, H., Leonard, D., \emph{et al.}\ 2012, \emph{Astrophys.\ J.}\ \textbf{752}, L22 (5pp).

\noindent Li, Y.~P., Gan, W.~Q.\ 2006, \emph{Astrophys.\ J.}\ \textbf{644}, L97--L100.

\noindent Liu, W., Ofman, L., Nitta, N.~V., Aschwanden, M.~J., Schrijver, C.~J., Title,\\
\hspace*{3mm} A.~M., Tarbell, T.~D.\ 2012, \emph{Astrophys.\ J.}\ \textbf{753}, 52 (17pp).

\noindent Liu, Y.\ 2008, \emph{Solar Phys.}\ \textbf{249}, 75--84.

\noindent Liu, Y., Luhmann, J.~G., Lin, R.~P., Bale, S.~D., Vourlidas, A., Petrie,\\
\hspace*{3mm} G.~J.~D.\ 2009, \emph{Astrophys.\ J.}\ \textbf{698}, L51--L55.

\noindent Lundquist, S.\ 1951, \emph{Phys.\ Rev.}\ \textbf{83}, 307--311.

\noindent Madjarska, M.~S.\ 2011, \emph{A\&A\/} \textbf{526}, A19 (24pp).

\noindent Mathioudakis, M., Jess, D.~B., Erd\'elyi, R.\ 2013, \emph{Space Sci.\ Rev.}\ \textbf{175}, 1--27.

\noindent McIntosh, S.~W., De Pontieu, B., Carlsson, M., Hansteen, V., Boerner, P., \\
\hspace*{3mm} Goossens, M.\ 2011, \emph{Nature\/} \textbf{475}, 477--480.

\noindent Morton, R.~J., Srivastava, A.~K., Erd\'elyi, R.\ 2012, \emph{A\&A\/} \textbf{542}, A70 (9pp).

\noindent M\"{o}stl, U.~V., Temmer, M., Veronig, A.~M.\ 2013, \emph{Astrophys.\ J.}\ \textbf{766}, L12 \\
\hspace*{3mm} (6pp).

\noindent Murawski, K., Solov'ev, A.~A., Kra\'{s}kiewicz, J., Srivastava, A.~K.\ 2015, \emph{A\&A}, \\
\hspace*{3mm} in press.

\noindent Murawski, K., Srivastava, A.~K., Zaqarashvili, T.~V.\ 2011, \emph{A\&A\/} \textbf{535}, A58 \\
\hspace*{3mm} (9pp).

\noindent Nakariakov, V.~M., Ofman, L., Deluca, E.~E., Roberts,~B., Davila, J.~M.\ \\
\hspace*{3mm} 1999, \emph{Science\/} \textbf{285}, 862--864.

\noindent Nakariakov, V.~M., Verwichte, E.\ 2005, \emph{Living Rev.\ Solar Phys.}\ \textbf{2}, 3--65.

\noindent Newton, H.~W.\ 1942, \emph{MNRAS\/} \textbf{102}, 108--109.

\noindent Nistic\`o, G., Nakariakov, V.~M., Verwichte, E.\ 2013, \emph{A\&A\/} \textbf{552}, A57 (6pp).

\noindent Nishizuka, N., Shimizu, M., Nakamura, T., Otsuji, K., Okamoto, T.~J., \\
\hspace*{3mm} Katsukawa, Y., Shibata, K.\ 2008, \emph{Astrophys.\ J.}\ \textbf{683}, L83--L86.

\noindent Ofman, L., Thompson, B.~J.\ 2011, \emph{Astrophys.\ J.}\ \textbf{734}, L11 (5pp).

\noindent O'Shea, E., Srivastava, A.~K., Doyle, J.~G., Banerjee, D.\ 2007, \emph{A\&A\/} \textbf{473},
\hspace*{3mm} L13--L16.

\noindent Panasenco, O., Martin, S.~F., Velli, M.\ 2014, \emph{Solar Phys.}\ \textbf{289}, 603--622.

\noindent Pariat, E., Antiochos, S.~K., DeVore, C.~R.\ 2009, \emph{Astrophys.\ J.}\ \textbf{691}, 61--74.

\noindent Patsourakos, S., Vourlidas, A., Stenborg, G.\ 2013, \emph{Astrophys.\ J.}\ \textbf{764}, 125 \\
\hspace*{3mm} (13pp).

\noindent Pereira, T.~M.~D., De Pontieu, B., Carlsson, M.\ 2013, \emph{Astrophys.\ J.}\ \textbf{764}, 69 \\
\hspace*{3mm} (5pp).

\noindent Pesnell, D.~W., Thompson, B.~J., Chamberlin, P.~C.\ 2012, \emph{Solar Phys.}\ \textbf{275}, \\
\hspace*{3mm} 3--15.

\noindent Popescu, M.~D., Xia, L.~D., Banerjee, D., Doyle, J.~G.\ 2007, \emph{Adv.\ Space Res.}\ \\
\hspace*{3mm} \textbf{40}, 1021--1025.

\noindent Priest, E.~R.\ 2014, \emph{Magnetohydrodynamics of the Sun}, Cambridge University \\
\hspace*{3mm} Press, New York.

\noindent Roberts, B., Webb, A.~R.\ 1978, \emph{Solar Phys.}\ \textbf{56}, 5--35.

\noindent Roberts, B.\ 1979, \emph{Solar Phys.}\ \textbf{61}, 23--34.

\noindent Roberts, B.\ 1981, in: \emph{The Physics of Sunspots}, J.~H.~Thomas and L.~E.~Cram \\
\hspace*{3mm} (eds), Sunspot, Sacramento Peak Observatory, New Mexico, pp.\ 369--383.

\noindent Roberts, B.\ 1991, \emph{Geophys.\ \& Astrophys.\ Fluid Dynamics\/} \textbf{62}, 83--100.

\noindent Roberts, B., Ulmschneider, P.\ 1997, in: \emph{Solar and Heliospheric Plasma \\
\hspace*{3mm} Physics}, G.~M.~Simett, C.~E.~Alissandrakis, and L.~Vlahos (eds), Springer,\\
\hspace*{3mm} Berlin, pp.~75--101.

\noindent Roberts, P.~H.\ 1956, \emph{Astrophys.\ J.}\ \textbf{124}, 430--442.

\noindent Ruderman, M.~S.\ 2007, \emph{Solar Phys.}\ \textbf{246}, 119--131.

\noindent Ruderman, M.~S., Erd\'elyi, R.\ 2009, \emph{Space Sci. Rev.}\ \textbf{149}, 199--228.

\noindent Savcheva, A., Cirtain, J., Deluca, E.~E., \emph{et al.}\ 2007, \emph{Publ.\ Astron.\ Soc.\ Japan\/} \\
\hspace*{3mm} \textbf{59}, 771--778.

\noindent Schmieder, B., D\'emoulin, P., Aulanier, G.\ 2013, \emph{Adv.\ Space Res.}\ \textbf{51},\\
\hspace*{3mm} 1967--1980.

\noindent Secchi, P.~A.\ 1877, \emph{Le Soleil}, 2nd edition, Part II, Gauthiers-Villars, Paris.

\noindent Shen, Y., Liu, Y., Su, J., Ibrahim, A.\ 2011, \emph{Astrophys.\ J.}\ \textbf{735}, L43 (5pp).

\noindent Shibata, K., Ishido, Y., Acton, L.~W., Strong, K.~T., Hirayama, T., Uchida,\\
\hspace*{3mm} Y.\ 1992, \emph{Publ.\ Astron.\ Soc.\ Japan\/} \textbf{44}, L173--L179.

\noindent Shimojo, M., Shibata, K.\ 2000, \emph{Astrophys.\ J.}\ \textbf{542}, 1100--1108.

\noindent Skogsrud, H., Rouppe van der Voort, L., De Pontieu, B.\ 2014, \emph{Astrophys.\ J.}\ \\
\hspace*{3mm} \textbf{795}, L23 (6pp).

\noindent Solanki, S.~K.\ 1993, \emph{Space Sci.\ Rev.}\ \textbf{63}, 1--188.

\noindent Soler R., Terradas, J., Oliver, R., Ballester, J.~L., Goossens, M.\ 2010, \\
\hspace*{3mm} \emph{Astrophys.\ J.}\ \textbf{712}, 875--882.

\noindent Soler R., D\'{i}az, A.~J., Ballester, J.~L., Goossens, M.\ 2012, \emph{Astrophys.\ J.}\ \textbf{749},\\
\hspace*{3mm} 163 (12pp).

\noindent Song, H.~Q., Chen, Y., Ye, D.~D., Han, G.~Q., Du, G.~H., Li,~G., Zhang,~J.,\\
\hspace*{3mm} Hu, Q.\ 2013, \emph{Astrophys.\ J.}\ \textbf{773}, 129 (10pp).

\noindent Song, H.~Q., Zhang, J., Chen, Y., Cheng, X.\ 2014, \emph{Astrophys.\ J.}\ \textbf{792}, L40 \\
\hspace*{3mm} (6pp).

\noindent Spruit, H.~C.\ 1981, \emph{A\&A\/} \textbf{98}, 155--160.

\noindent Spruit, H.~C.\ 1982, \emph{Solar Phys.}\ \textbf{75}, 3--17.

\noindent Srivastava, A.~K., Goossens, M.\ 2013, \emph{Astrophys.\ J.}\ \textbf{777}, 17 (9pp).

\noindent Srivastava, A.~K., Murawski, K.\ 2011, \emph{A\&A\/} \textbf{534}, A62 (7pp).

\noindent Su, Y., van Ballegooijen, A.\ 2012, \emph{Astrophys.\ J.}\ \textbf{757}, 168 (17pp).

\noindent Su, Y., van Ballegooijen, A.\ 2013, \emph{Astrophys.\ J.}\ \textbf{764}, 91 (17pp).

\noindent Taroyan, Yu., Ruderman, M.~S.\ 2011, \emph{Space Sci.\ Rev.}\ \textbf{158}, 505--523.

\noindent Tavabi, E., Koutchmy, S., Ajabshirizadeh, A.\ 2011, \emph{New Astron.}\ \textbf{16},\\
\hspace*{3mm} 296--305.

\noindent Temmer, M., Veronig, A.~M., Kontar, E.~P., Krucker, S., Vr\v{s}nak, B.\ 2010, \\
\hspace*{3mm} \emph{Astrophys.\ J.}\ \textbf{712}, 1410--1420.

\noindent Tian, H., McIntosh, S.~W., Wang, T., Ofman, L., De Pontieu, B., Innes, \\
\hspace*{3mm} D.~E., Peter, H.\ 2012, \emph{Astrophys.\ J.}\ \textbf{759}, 144 (17pp).

\noindent Tomczyk, S., McIntosh, S.~W., Keil, S.~L., Judge P.~G., Schad, T., Seeley,\\
\hspace*{3mm} D.~H., Edmondson, J.\ 2007, \emph{Science\/} \textbf{317}, 1192--1196.

\noindent T\"{o}r\"{o}k, T., Kliem, B.\ 2005, \emph{Astrophys.\ J.}\ \textbf{630}, L97--L100.

\noindent Trehan, S.~K., Reid, W.~H.\ 1958, \emph{Astrophys.\ J.}\ \textbf{127}, 454--458.

\noindent Tsiropoula, G., Tziotziou, K., Kontogiannis, I., Madjarska, M.~S., Doyle,\\
\hspace*{3mm} J.~G., Suematsu, Y.\ 2012, \emph{Space Sci.\ Rev.}\ \textbf{169}, 181--244.

\noindent Van Doorsselaere, T., Nakariakov, V.~M., Verwichte, E.\ 2008, \emph{Astrophys.\ J.}\ \\
\hspace*{3mm} \textbf{676}, L73--L75.

\noindent Vasheghani Farahani, S., Van Doorsselaere, T., Verwichte, E., Nakariakov,\\
\hspace*{3mm} V.~M.\ 2009, \emph{A\&A\/} \textbf{498}, L29--L32.

\noindent Vemareddy, P, Zhang, J.\ 2014, \emph{Astrophys.\ J.}\ \textbf{797}, 80 (12pp).

\noindent Verwichte E., Aschwanden, M.~J., Van Doorsselaere, T., Foullon, C.,\\
\hspace*{3mm} Nakariakov, V.~M.\ 2009, \emph{Astrophys.\ J.}\ \textbf{698}, 397--404.

\noindent Verwichte E., Van Doorsselaere T., Foullon C., White, R.~S.\ 2013, \emph{Astrophys.\ \\
\hspace*{3mm} J.}\ \textbf{767}, 16 (7pp).

\noindent Vourlidas, A.\ 2014, \emph{Plasma Phys.\ Control.\ Fusion\/} \textbf{56}, 064001 (6pp).

\noindent Vourlidas, A., Lynch, B.~J., Howard, R.~A., Li, Y.\ 2013, \emph{Solar Phys.}\ \textbf{284},\\
\hspace*{3mm} 179--201.

\noindent Wang, T., Ofman, L., Davila, J.~M., Su, Y.\ 2012, \emph{Astrophys.\ J.}\ \textbf{751}, L27 \\
\hspace*{3mm} (6pp).

\noindent Wang, T.~J., Solanki, S.~K.\ 2004, \emph{A\&A\/} \textbf{421}, L33--L36.

\noindent Webb, D.~F., Howard, T.~A.\ 2012, \emph{Living Rev.\ Solar Phys.}\ \textbf{9}, 3--83.

\noindent Wedemeyer-B\"{o}hm, S., Scullion, E., Steiner, O., \emph{et al.}\ 2012, \emph{Nature\/} \textbf{486},\\
\hspace*{3mm} 505--508.

\noindent White, R.~S., Verwichte E., Foullon, C.\ 2012, \emph{A\&A\/} \textbf{545}, A129 (10pp).

\noindent Yang, L., Zhang, J., Liu, W., Li, T., Shen, Y.\ 2013, \emph{Astrophys.\ J.}\ \textbf{775}, 39 \\
\hspace*{3mm} (12pp).

\noindent Yokoyama, T., Shibata, K.\ 1995, \emph{Nature\/} \textbf{375}, 42--44.

\noindent Yurchyshyn, V., Abramenko, V., Kosovichev, A., Goode, P.\ 2014, \emph{Astrophys.\ \\
\hspace*{3mm} J.}\ \textbf{787}, 58 (7pp).

\noindent Zaqarashvili, T.~V., D\'{i}az, A.~J., Oliver, R., Ballester, J.~L.\ 2010, \emph{A\&A\/} \textbf{516},\\
\hspace*{3mm} A84 (8pp).

\noindent Zaqarashvili, T.~V.\ 2011, in: \emph{AIP Conf.\ Proc.}\ \textbf{1356}, I.~Zhelyazkov and \\
\hspace*{3mm} T.~Mishonov (eds), AIP Publishing LLC, Melville, New York,\\
\hspace*{3mm} pp.~106--116.

\noindent Zaqarashvili, T.~V., V\"{o}r\"{o}s, Z., Zhelyazkov, I.\ 2014a, \emph{A\&A\/} \textbf{561}, A62 (7pp).

\noindent Zaqarashvili, T. V., V\"{o}r\"{o}s, Z., Narita, Y., Bruno, R.\ 2014b, \emph{Astrophys.\ J.}\ \\
\hspace*{3mm} \textbf{783}, L19 (4pp).

\noindent Zhang, J., Cheng, X., Ding, M.\ 2012, \emph{Nature Communications} \textbf{3}, 747 (5pp).

\noindent Zhang, J., Liu, Y.\ 2011, \emph{Astrophys.\ J.}\ \textbf{741}, L7 (5pp).

\noindent Zhang, J., Richardson, I.~G., Webb, D.~F., \emph{et al.}\ 2007, \emph{J.\ Geophys.\ Res.: \\
\hspace*{3mm} Space Phys.}\ \textbf{112}, A10102 (19pp).

\noindent Zhang, M., Low, B.~C.\ 2005, \emph{Ann.\ Rev.\ Astron.\ Astrophys.}\ \textbf{43}, 103--137.

\noindent Zhang, Q.~M., Ji, H.~S.\ 2014, \emph{A\&A\/} \textbf{561}, A134 (7pp).

\noindent Zhelyazkov, I.\ 2012a, \emph{A\&A\/} \textbf{537}, A124 (8pp).

\noindent Zhelyazkov, I., Zaqarashvili, T.~V.\ 2012, \emph{A\&A\/} \textbf{547}, A14 (11pp).

\noindent Zhelyazkov, I.\ 2012b, in: \emph{Topics in Magnetohydrodynamics}, L.~Zheng (ed),\\
\hspace*{3mm} InTech Publishing, Rijeka, Chap.~6, pp.~135--166.

\noindent Zhelyazkov, I.\ 2013, in: \emph{AIP Conf.\ Proc.}\ \textbf{1551}, I.~Zhelyazkov and \\
\hspace*{3mm} T.~Mishonov (eds), AIP Publishing LLC, Melville, New York, pp.~150--164.

\noindent Zhelyazkov, I., Chandra, R.\ 2014, \emph{Compt.\ rend.\ Acad.\ bulg. Sci.}\ \textbf{67},\\
\hspace*{3mm} 1145--1152.

\noindent Zhelyazkov, I., Chandra, R., Sivastava, A.~K., Mishonov, T.\ 2014, \emph{Astrophys.\ \\
\hspace*{3mm} Space Sci.}\ {\color{blue}{http://dx.doi.org/10.1007/s10509-014-2215-1}} (10pp).

\noindent Zhelyazkov, I., Zaqarashvili, T.~V., Chandra, R.\ 2015, \emph{A\&A\/} \textbf{574}, A55 (7pp).

\noindent Zheng, R., Jiang, Y., Yang, J., Bi, Y., Hong, J., Yang, B., Yang, D.\ 2013, \\
\hspace*{3mm} \emph{Astrophys.\ J.}\ \textbf{764}, 70 (7pp).

\end{document}